\documentclass[lettersize,journal]{IEEEtran}
\IEEEoverridecommandlockouts
\usepackage{amsmath,amsfonts,amsthm,amssymb}
\usepackage{array}
\usepackage{textcomp}
\usepackage{subfigure}
\usepackage{float}
\usepackage{enumerate}
\usepackage{tabularx}
\usepackage{stfloats}
\usepackage{url}
\usepackage{hyperref}
\usepackage{xcolor,soul,framed} 
\usepackage{pstricks}
\usepackage{verbatim}
\usepackage{graphicx}
\usepackage{epstopdf}
\usepackage{booktabs}
\usepackage{cite}
\usepackage[T1]{fontenc}
\usepackage{tikz}
\usepackage{algorithm}
\usepackage{algpseudocode}

\allowdisplaybreaks

\newenvironment{solidproof}
  {\proof}
  {\endproof}
\begin{document}
\newtheorem{Remark}{Remark}
\newtheorem{remark}{Remark}
\newtheorem{thm}{Theorem}
\newtheorem{lemma}{Lemma}
\renewcommand{\algorithmicrequire}{\textbf{Input:}} 
\renewcommand{\algorithmicensure}{\textbf{Output:}}

\title{Microwave Linear Analog Computers Aided Multiuser Communication: General Impedance Matching and Precoding Optimization\\
}
\author{Rujing~Xiong,~\IEEEmembership{Member,~IEEE,}
        Jie~Xu,~\IEEEmembership{Fellow,~IEEE,}
        and Rui~Zhang,~\IEEEmembership{Fellow,~IEEE}
\thanks{R.~Xiong  is with the School of Science and Engineering, The Chinese University of Hong Kong (Shenzhen), Shenzhen 518172, China (e-mail: rujingxiong@cuhk.edu.cn).

J.~Xu is with the School of Science and Engineering and Shenzhen Future Network of Intelligence Institute (FNii), The Chinese University of Hong Kong (Shenzhen), Shenzhen 518172, China (e-mail: xujie@cuhk.edu.cn).

R.~Zhang is with the Department of Electrical and Computer Engineering, National University of Singapore, Singapore 117583 (e-mail: elezhang@nus.edu.sg).}
\thanks{(\textit{Corresponding authors: Jie Xu; Rui Zhang})}

}
\maketitle

\begin{abstract}
Microwave linear analog computers (MiLACs) have recently emerged as a hardware-efficient solution for implementing multi-antenna communication systems. Unlike existing MiLAC designs based on the ideal assumption of perfect impedance matching (PIM) with reflection-free transmission, this paper investigates MiLAC-aided precoding optimization under a general impedance matching (GIM) model, which enables more flexible precoder design at the cost of a potential reduction in radiated power. Specifically, we consider a downlink multiuser multiple-input single-output (MISO) communication system and aim to maximize the system sum rate by optimizing the MiLAC-enabled transmit precoding subject to physical circuit constraints. The formulated problem is challenging to solve due to the intricate coupling between the precoding and impedance parameters. To address this challenge, we first develop a singular value decomposition (SVD)-based parametric search framework for small or medium size systems. This framework exploits the feasible precoder structure and explicitly captures the tradeoff between power radiation efficiency and precoder design flexibility. We then propose a unified algorithm for solving the optimization problem based on the projected weighted minimum mean-square error (WMMSE) principle for arbitrary size systems with GIM- or PIM-based MiLAC precoding. Simulation results demonstrate that the GIM-based MiLAC design consistently outperforms its PIM counterpart as a special realization, especially in interference-limited scenarios, by allowing a moderate reduction in radiated power in exchange for additional precoder design flexibility and more effective interference mitigation. It is also shown that GIM-based MiLAC design achieves performance close to that of the baseline fully digital precoding system.
\end{abstract}


\section{Introduction}
Future sixth-generation (6G) wireless networks are expected to support unprecedented requirements in terms of data rate, connectivity density, latency, reliability, and energy efficiency. To meet these demands, multi-antenna technologies, especially ultra-massive/gigantic multiple-input multiple-output (MIMO), extremely large-scale antenna arrays, and distributed multi-antenna architectures~\cite{Emil2025enabling6g,carvalho2020nonstationarities,xu2025distributed}, have become central components of 6G wireless systems, as they provide abundant spatial degrees of freedom for spatial multiplexing, diversity enhancement, precoder design, and interference mitigation. In particular, gigantic MIMO systems may scale the number of antennas to hundreds or even thousands, far beyond conventional massive MIMO architectures deployed in fifth-generation (5G) systems~\cite{larsson2014massive,emil2016massive}.
However, as the number of antennas continues to grow, conventional fully digital precoders face severe implementation bottlenecks. Although fully digital design offers the highest precoder design flexibility and spatial multiplexing capability, it typically requires dedicated radio-frequency (RF) chain and high-resolution analog-to-digital converter (ADC)/digital-to-analog converter (DAC) for each antenna element. Moreover, as the antenna dimension increases, the volume of baseband data to be processed in real time grows rapidly, requiring high-dimensional symbol-level matrix operations. These requirements lead to prohibitive hardware complexity, power consumption, and cost.

Recently, microwave linear analog computers (MiLACs) have emerged as a promising hardware-efficient architecture for multi-antenna communication systems~\cite{nerini2025analogI,nerini2025analog}. A MiLAC is a reconfigurable multiport microwave network composed of tunable impedance components, capable of directly implementing linear transformations in the electromagnetic domain. It has been shown in the computing field that MiLACs can efficiently realize fundamental linear operations, such as linear minimum mean-square error (LMMSE) estimation and matrix inversion, with substantially reduced computational complexity compared with conventional digital implementations~\cite{nerini2025analogI}. 
When applied to wireless communications, MiLACs enable analog-domain precoding and combining by processing RF signals through reconfigurable microwave networks. This reduces the number of RF chains, alleviates high-dimensional baseband signal processing, eliminates per-symbol digital matrix-vector multiplications, and facilitates the use of low-resolution data converters~\cite{nerini2025analog}. 

In the literature, a handful of prior works have investigated MiLAC-aided communications and demonstrated their potential. For point-to-point MIMO systems, lossless and reciprocal MiLAC precoder was shown in~\cite{nerini2025capacity} to achieve the same data rate as fully digital precoding scheme with the same number of data streams. Reduced-complexity MiLAC architectures were further developed in~\cite{nerini2025reduced} to preserve the capacity-achieving property while significantly reducing the number of tunable circuit components. For multiuser multiple-input single-output (MISO) systems with $K$ users, it was shown in~\cite{foad2016hybrid,wu2026microwavelinearanalogcomputer} that hybrid digital–MiLAC design with $K$ RF chains can achieve the same flexibility as fully digital precoder design, thereby halving the RF-chain requirement compared with conventional hybrid precoding, which requires $2K$ RF chains to achieve full flexibility. More recently, the authors in~\cite{zhou2026twolayer} derived a scattering-matrix construction method that maps arbitrary fully digital precoders to a two-layer MiLAC architecture with intermediate power allocation, achieving the same rate performance as fully digital precoding system while using only $K$ low-resolution RF chains. In addition, hybrid digital–MiLAC precoder has been extended to wideband multiuser MISO orthogonal frequency-division multiplexing (OFDM) systems, where a frequency-flat MiLAC is shared across subcarriers and a low-dimensional digital precoder is used to handle frequency selectivity~\cite{peng2026hybriddigital}. Beyond precoding, MiLACs have also been employed for analog-domain channel estimation, where least-squares (LS) and minimum mean-square error (MMSE) estimation were implemented with the same performance as their digital counterparts, while reducing online digital computation, RF-chain usage, data-converter resolution requirements, and peak-to-average power ratio (PAPR)~\cite{zhang2026channeles}.

Despite these advances, incorporating MiLACs into transmit precoding introduces new challenges that are fundamentally tied to their microwave-network implementation. A physically consistent MiLAC must satisfy practical circuit constraints such as passivity, losslessness, and reciprocity~\cite{nerini2026physics}. Most existing studies focus on idealized perfect impedance matching (PIM) assumption or reflection-free configuration, where no RF-side reflection occurs and the RF-to-antenna transmission efficiency is maximized. With PIM, the induced forward transmission matrix becomes semi-unitary, resulting in unit transmission efficiency and a highly structured precoder~\cite{nerini2025capacity,fang2026performance}. Although this PIM structure can achieve the same data rate as its fully digital precoding counterpart in point-to-point MIMO systems~\cite{nerini2025capacity}, it generally fails to provide full digital precoder design flexibility in multiuser MISO systems with MiLAC-only precoder, especially when the user channels become highly non-orthogonal~\cite{wu2026microwavelinearanalogcomputer,fang2026performance}. Moreover, maintaining PIM in practical microwave networks is challenging due to frequency selectivity, hardware nonidealities, load variations, port coupling, and antenna mutual coupling~\cite{fano1950theoretical,nemati2009design,alibakhshikenari2021optimum,nerini2024universal}.

Motivated by these limitations, this paper departs from the ideal PIM assumption and investigates MiLAC-aided multiuser communications under a general impedance matching (GIM) model. From a scattering-matrix perspective, PIM enforces zero RF-side reflection and restricts the induced precoder to a semi-unitary structure, whereas GIM allows non-zero reflection at the RF interface. This relaxation enlarges the feasible precoding space and provides additional degrees of freedom for multiuser interference mitigation, at the cost of a reduced MiLAC transmission coefficient or effective radiated power. Therefore, GIM introduces a fundamental tradeoff between power radiation efficiency and precoder design flexibility, which can be exploited for performance enhancement. However, how to characterize this tradeoff and leverage it for sum-rate improvement in multiuser scenarios remains unexplored.

To fill this gap, we investigate in this paper a MiLAC-aided downlink multiuser MISO communication system under the GIM model. The main contributions of this paper are summarized as follows:

\begin{itemize}
    \item First, we establish a physics-compliant framework for MiLAC-aided multiuser MISO precoding optimization, which provides closed-form expressions explicitly linking the susceptance matrix, scattering matrix, induced precoding matrix, and MiLAC transmission coefficient. Based on this framework, we characterize the structural difference between GIM and PIM in the precoder domain. Specifically, we show that the MiLAC-induced precoder is inherently constrained by the RF-to-antenna transmission structure of the scattering matrix. Under the GIM model, the induced precoder satisfies a contraction constraint, allowing singular-value attenuation and non-zero RF-side reflection. In contrast, PIM enforces a semi-unitary precoder structure through the zero-reflection condition. This characterization explicitly reveals the tradeoff between the MiLAC transmission coefficient and precoder design flexibility, showing that a radiated-power loss can provide additional degrees of freedom for precoder design and interference mitigation.

\item Next, we aim to maximize the system sum rate by optimizing the MiLAC-enabled transmit precoder subject to physical circuit constraints. We first develop a singular value decomposition (SVD)-based parametric search framework to characterize the feasible precoding structures under the GIM or PIM model, and to provide benchmark solutions for small or medium size systems. We then propose a unified algorithm based on the projected weighted minimum mean-square error (WMMSE) for arbitrary size systems with MiLAC precoding. For both GIM- and PIM-based designs, the proposed algorithm employs appropriate projections onto their respective feasible sets.

\item Finally, simulation results demonstrate that, by allowing a moderate reduction in radiated power in exchange for additional precoder design flexibility, the GIM-based MiLAC design consistently outperforms its PIM counterpart, especially in interference-limited scenarios. It is also shown that GIM-based MiLAC design achieves performance close to that of the fully digital precoding benchmark.

\end{itemize}

The rest of this paper is organized as follows. 
Section~\ref{Section2} introduces the MiLAC-aided downlink multiuser MISO communication model and formulates the sum-rate maximization problem under the GIM and PIM models. 
Section~\ref{Section3} develops the SVD-based feasible-set characterization and search framework for small or medium size systems. 
Section~\ref{Section4} proposes a unified projected WMMSE-based algorithm for arbitrary size systems with MiLAC precoding optimization.
Section~\ref{Section5} presents numerical results to evaluate the power-radiation-efficiency and precoder-design-flexibility tradeoff and to compare MiLAC with fully digital benchmarks. 
Finally, Section~\ref{Section6} concludes this paper.

\textit{Notations:}
Scalars, vectors, and matrices are denoted by lowercase letters, bold lowercase letters, and bold uppercase letters, respectively. 
For a scalar, vector, or matrix, $(\cdot)^{*}$ denotes the complex conjugate. 
For a matrix $\mathbf{A}$, $\mathbf{A}^{T}$, $\mathbf{A}^{H}$, $\mathbf{A}^{-1}$, $\mathrm{tr}(\mathbf{A})$, and $\|\mathbf{A}\|_{F}$ denote its transpose, Hermitian transpose, inverse, trace, and Frobenius norm, respectively. 
The notation $\|\mathbf{a}\|_{2}$ denotes the Euclidean norm of vector $\mathbf{a}$. 
The notation $\mathbf{A}\preceq\mathbf{B}$ means that $\mathbf{B}-\mathbf{A}$ is positive semidefinite. 
The identity matrix is denoted by $\mathbf{I}$, and the all-zero matrix is denoted by $\mathbf{0}$ with proper dimensions. 
The operator $\operatorname{diag}(\cdot)$ forms a diagonal matrix from its arguments, and $\operatorname{Re}\{\cdot\}$ extracts the real part of a complex-valued argument. 
The expectation operator is denoted by $\mathbb{E}[\cdot]$, and $\mathcal{CN}(0,\sigma^2)$ denotes the circularly symmetric complex Gaussian (CSCG) distribution with zero mean and variance $\sigma^2$. 
The imaginary unit is denoted by $j$.
The sets of real and complex matrices of size $M\times N$ are denoted by $\mathbb{R}^{M\times N}$ and $\mathbb{C}^{M\times N}$, respectively. Unless otherwise specified, $\log(\cdot)$ denotes the natural logarithm.

\section{System Model and Problem Formulation}\label{Section2}
In this section, we first introduce the downlink multiuser MISO communication model and then describe a physically realizable MiLAC as the precoder. Based on this model, we formulate the MiLAC sum-rate maximization problem. 

\begin{figure}[tbp]
\centering
\includegraphics[width=1\linewidth]{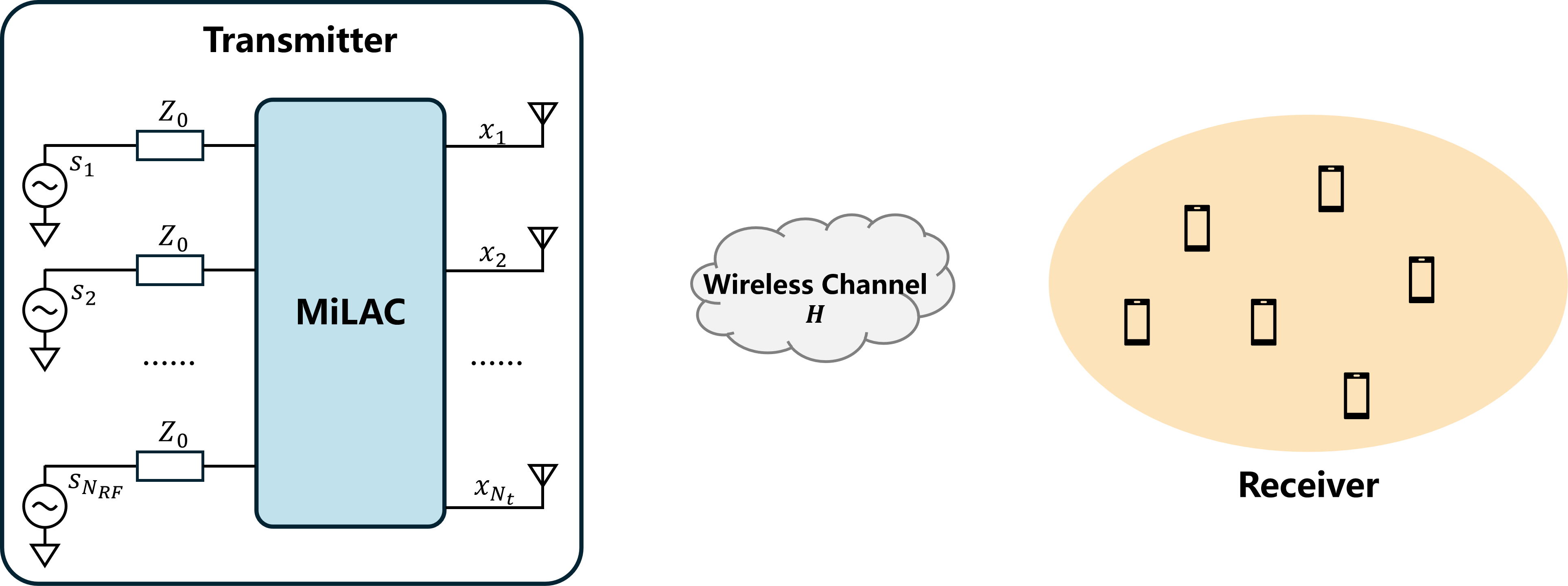}
\caption{MiLAC-aided downlink multiuser communications.}
\label{F0}
\end{figure}

\subsection{Downlink Multiuser MISO Communication}

We consider a narrowband downlink multiuser MISO system, in which a base station (BS) serves $K$ single-antenna users simultaneously. The BS is equipped with $N_t$ transmit antennas and a reciprocal and lossless MiLAC with $N_{RF}$ RF input ports and $N_t$ antenna output ports, as shown in Fig.~\ref{F0}, with $N_t\geq N_{RF}$ in general. Since each user is assigned one independent data stream, the number of transmitted streams is $N_s=K$, and we set $N_{RF}=K$ to support these streams through the MiLAC architecture. Let
$
\mathbf{s}=[s_1,\ldots,s_K]^T\in\mathbb{C}^{K\times 1}
$ 
denote the transmitted symbol vector for the $K$ users, with normalized covariance
$
\mathbb{E}\left[\mathbf{s}\mathbf{s}^{H}\right]=\mathbf{I}_K
$.
Let $\mathbf{F}=[\mathbf{f}_1,\ldots,\mathbf{f}_K]\in\mathbb{C}^{N_t\times K}$ denote a generic precoding matrix, where $\mathbf{f}_k$ is the transmit precoding vector for user $k$. The physical realization of $\mathbf{F}$ through the MiLAC network will be specified in the next subsection. 
The transmitted signal is modeled as
\begin{equation}
\label{EQ:GenericTxSignal}
\mathbf{x}
=
\sqrt{\frac{P_t}{K}}\mathbf{F}\mathbf{s},
\end{equation}
where $P_t$ denotes the total transmit-power budget. Let
$
\mathbf{H}
=
[\mathbf{h}_1,\mathbf{h}_2,\ldots,\mathbf{h}_K]
\in\mathbb{C}^{N_t\times K}
$
denote the aggregate downlink channel matrix, where $\mathbf{h}_k\in\mathbb{C}^{N_t\times 1}$ is the channel vector from the BS transmitter to user $k$. The received signal at user $k$ is given by
\begin{equation}
\label{EQ:ReceivedSignalGeneric}
y_k
=
\sqrt{\frac{P_t}{K}}
\mathbf{h}_k^{H}\mathbf{f}_k s_k
+
\sqrt{\frac{P_t}{K}}
\sum_{i\neq k}
\mathbf{h}_k^{H}\mathbf{f}_i s_i
+
n_k,
\end{equation}
where $n_k\sim\mathcal{CN}(0,\sigma_k^2)$ denotes the additive white Gaussian noise at user $k$. Accordingly, the received signal-to-interference-plus-noise ratio (SINR) at user $k$ is given by
\begin{equation}
\label{EQ:SINRGeneric}
\gamma_k(\mathbf{F})
=
\frac{
\frac{P_t}{K}
\left|
\mathbf{h}_k^{H}\mathbf{f}_k
\right|^2
}{
\frac{P_t}{K}
\sum_{i\neq k}
\left|
\mathbf{h}_k^{H}\mathbf{f}_i
\right|^2
+
\sigma_k^2
}.
\end{equation}
The achievable rate of user $k$ in bit-per-second-per-Hertz (bit/s/Hz) is
$
R_k(\mathbf{F})
=
\log_2\left(1+\gamma_k(\mathbf{F})\right)$, 
and the system sum-rate is
\begin{equation}
\label{EQ:SumRateGeneric}
R_{\mathrm{sum}}(\mathbf{F})
=
\sum_{k=1}^{K} R_k(\mathbf{F})  =\sum_{k=1}^K
\log_2\left(1+\gamma_k(\mathbf{F})\right).
\end{equation}

\subsection{Physics-Compliant MiLAC Precoding}
We introduce the realization of the precoding matrix $\mathbf{F}$ via the MiLAC in this subsection. Specifically, when the MiLAC is employed for transmit precoding, the RF-to-antenna transmission block of its scattering matrix serves as the equivalent baseband precoding matrix, as will be detailed next. To support the $K$ data
streams, we set the number of RF input ports as 
$
N_{RF}=K.
$
Let $N=N_{RF}+N_t$ denote the total number of ports of the MiLAC and $Z_0$ denote the reference impedance. The scattering matrix is determined by the susceptance matrix $\mathbf{B}$ through the Cayley transform~\cite{nerini2025capacity}, given by
\begin{equation}
\label{EQ:Cayley}
\boldsymbol{\Theta}(\mathbf{B})
=
\left(\mathbf{I}-jZ_0\mathbf{B}\right)
\left(\mathbf{I}+jZ_0\mathbf{B}\right)^{-1},
\end{equation}
where
$\mathbf{B}=\mathbf{B}^{T}\in\mathbb{R}^{N\times N}$.
By separating the RF input ports and antenna output ports, $\boldsymbol{\Theta}(\mathbf{B})$  is partitioned as
\begin{equation}
\boldsymbol{\Theta}(\mathbf{B})
=
\begin{bmatrix}
\boldsymbol{\Theta}_{11}(\mathbf{B}) & \boldsymbol{\Theta}_{12}(\mathbf{B})\\
\boldsymbol{\Theta}_{21}(\mathbf{B}) & \boldsymbol{\Theta}_{22}(\mathbf{B})
\end{bmatrix}.
\end{equation}
Here, $\boldsymbol{\Theta}_{11}(\mathbf{B})\in\mathbb{C}^{N_{RF}\times N_{RF}}$ denotes the RF-side reflection block, $\boldsymbol{\Theta}_{21}(\mathbf{B})\in\mathbb{C}^{N_t\times N_{RF}}$ denotes the RF-to-antenna transmission block, $\boldsymbol{\Theta}_{12}(\mathbf{B})\in\mathbb{C}^{N_{RF}\times N_t}$ denotes the antenna-to-RF block, and
$\boldsymbol{\Theta}_{22}(\mathbf{B})\in\mathbb{C}^{N_t\times N_t}$ denotes the antenna-side reflection block. The scattering matrix satisfies~\cite{wu2026microwavelinearanalogcomputer}
\begin{equation}
\label{EQ:ThetaUnitarySymmetric}
\boldsymbol{\Theta}^{H}(\mathbf{B})\boldsymbol{\Theta}(\mathbf{B})=\mathbf{I},
\qquad
\boldsymbol{\Theta}^{T}(\mathbf{B})=\boldsymbol{\Theta}(\mathbf{B}).
\end{equation}

The RF-to-antenna transmission block directly defines the equivalent baseband precoding matrix\footnote{For notational simplicity, the constant scaling factor associated with the standard power-wave normalization is absorbed into the transmit-power definition~\cite{wu2026microwavelinearanalogcomputer}.}~\cite{nerini2025capacity}, i.e.,
\begin{equation}
\label{EQ:F_B}
\mathbf{F}(\mathbf{B})
\triangleq
\boldsymbol{\Theta}_{21}(\mathbf{B}),
\qquad
\mathbf{F}(\mathbf{B})\in\mathbb{C}^{N_t\times K}.
\end{equation}
The precoding matrix in~\eqref{EQ:GenericTxSignal} is specified as $\mathbf{F}=\mathbf{F}(\mathbf{B})$. Unlike fully digital precoding matrices, which are directly optimized over arbitrary complex-valued entries, the MiLAC-induced precoding matrix is generated through the structured physical mapping
$
\mathbf{B}
\mapsto
\boldsymbol{\Theta}(\mathbf{B})
\mapsto
\mathbf{F}(\mathbf{B}).
$
Therefore, the feasible set of MiLAC precoding optimization is inherently governed by practical constraints on the physical microwave networks.

As a fundamental property of the microwave network, impedance matching characterizes how efficiently incident RF signals are transferred from the RF input ports to the antenna output ports. In the MiLAC-aided transmitter, GIM allows a non-zero RF-side reflection, whereas PIM enforces a zero-reflection constraint at the RF ports.

With GIM, the MiLAC is required only to satisfy the lossless reciprocal condition $\mathbf{B}=\mathbf{B}^{T}\in\mathbb{R}^{N\times N}$, without imposing any additional constraint on the RF-side reflection block $\boldsymbol{\Theta}_{11}(\mathbf{B})$. From the unitarity of $\boldsymbol{\Theta}(\mathbf{B})$ in~\eqref{EQ:ThetaUnitarySymmetric}, we have
\begin{equation}
\label{EQ:ThetaEnergy}
\boldsymbol{\Theta}_{21}^{H}(\mathbf{B})
\boldsymbol{\Theta}_{21}(\mathbf{B})
=
\mathbf{I}_{N_{RF}}
-
\boldsymbol{\Theta}_{11}^{H}(\mathbf{B})
\boldsymbol{\Theta}_{11}(\mathbf{B})
\preceq
\mathbf{I}_{N_{RF}}.
\end{equation}
Equivalently, the induced precoding matrix satisfies
\begin{equation}
\label{EQ:FContraction}
\mathbf{F}^{H}(\mathbf{B})\mathbf{F}(\mathbf{B})
\preceq
\mathbf{I}_{N_{RF}}.
\end{equation}
This relation shows that non-zero RF-side reflection reduces the power transferred from the RF ports to the antenna ports.

To quantify this effect, and motivated by the power conservation relation of lossless scattering networks, we define the MiLAC transmission coefficient as
\begin{equation}
\label{EQ:EtaTx}
\eta_{\mathrm{tx}}(\mathbf{B})
=
\frac{\left\|\boldsymbol{\Theta}_{21}(\mathbf{B})\right\|_{F}^{2}}{N_{RF}}
=
1-
\frac{\left\|\boldsymbol{\Theta}_{11}(\mathbf{B})\right\|_{F}^{2}}{N_{RF}}.
\end{equation}
This parameter measures the RF-to-antenna transmission efficiency of the MiLAC. Under the GIM model, we generally have $\eta_{\mathrm{tx}}(\mathbf{B})\leq 1$.

Substituting $\mathbf{F}=\mathbf{F}(\mathbf{B})$ into~\eqref{EQ:GenericTxSignal}, the actual radiated power becomes
\begin{equation}
\label{EQ:Prad}
P_{\mathrm{rad}}(\mathbf{B})
=
\mathbb{E}\left[\|\mathbf{x}\|_2^2\right]
=
\frac{P_t}{K}
\mathrm{tr}
\left(
\mathbf{F}^{H}(\mathbf{B})\mathbf{F}(\mathbf{B})
\right)
=
P_t\eta_{\mathrm{tx}}(\mathbf{B}).
\end{equation}
Hence, GIM allows a broader class of realizable precoders but may reduce the radiated power due to non-zero RF-side reflection.

\begin{remark}
PIM is a special case of GIM obtained by imposing the additional zero-reflection constraint
\begin{equation}
\boldsymbol{\Theta}_{11}(\mathbf{B})=\mathbf{0}.
\end{equation}
Under this condition, \eqref{EQ:ThetaEnergy} reduces to
\begin{equation}\label{Condi:PIM}
\mathbf{F}^{H}(\mathbf{B})\mathbf{F}(\mathbf{B})
=
\boldsymbol{\Theta}_{21}^{H}(\mathbf{B})
\boldsymbol{\Theta}_{21}(\mathbf{B})
=
\mathbf{I}_{N_{RF}},
\end{equation}
and thus $\eta_{\mathrm{tx}}(\mathbf{B})=1$. Therefore, PIM eliminates RF-side reflection and guarantees a unit transmission coefficient. However, PIM also restricts the precoding matrix to a semi-unitary structure. In contrast, GIM may incur transmission loss due to non-zero reflection, but it provides a larger feasible precoding space. This reveals a fundamental tradeoff between MiLAC radiated power and precoder design flexibility.
\end{remark}

\subsection{Problem Formulation}

Based on the above model, the MiLAC sum-rate maximization problem under the GIM model is formulated as
\begin{equation}
\label{EQ:P1}
\begin{aligned}
\mathrm{(P1)}:\quad
\max_{\mathbf{B}}\quad
&
R_{\mathrm{sum}}\left(\mathbf{F}(\mathbf{B})\right)
\\
\mathrm{s.t.}\quad
&
\mathbf{B}=\mathbf{B}^{T},
\quad
\mathbf{B}\in\mathbb{R}^{N\times N},
\\
&
\boldsymbol{\Theta}(\mathbf{B})
=
\left(\mathbf{I}-jZ_0\mathbf{B}\right)
\left(\mathbf{I}+jZ_0\mathbf{B}\right)^{-1},
\\
&
\mathbf{F}(\mathbf{B})
=
\boldsymbol{\Theta}_{21}(\mathbf{B}).
\end{aligned}
\end{equation}
Problem (P1) captures the most general lossless reciprocal MiLAC design space considered in this work. Since no zero-reflection constraint is imposed, the problem allows the optimizer to balance the MiLAC radiated power and the precoder design flexibility. In particular, a non-zero $\boldsymbol{\Theta}_{11}(\mathbf{B})$ may reduce the radiated power through $\eta_{\mathrm{tx}}(\mathbf{B})$, but it can also enlarge the feasible precoding space to improve interference shaping.

\begin{remark}
Let the feasible set of the susceptance matrix under the GIM and PIM models be defined as
$
\mathcal{B}_{\mathrm{GIM}}
\triangleq
\left\{
\mathbf{B}\in\mathbb{R}^{N\times N}
\,\middle|\,
\mathbf{B}=\mathbf{B}^{T}
\right\}$, 
and
$
\mathcal{B}_{\mathrm{PIM}}
\triangleq
\left\{
\mathbf{B}\in\mathcal{B}_{\mathrm{GIM}}
\,\middle|\,
\boldsymbol{\Theta}_{11}(\mathbf{B})=\mathbf{0}
\right\}$.
Since PIM imposes an additional zero-reflection constraint on top of the GIM model, it follows that
\begin{equation}
\label{EQ:IN}
\mathcal{B}_{\mathrm{PIM}}
\subseteq
\mathcal{B}_{\mathrm{GIM}}.
\end{equation}
Define the globally optimal sum-rates under GIM and PIM as
$$
R_{\mathrm{sum}}^{\star,\mathrm{GIM}}
\triangleq
\max_{\mathbf{B}\in\mathcal{B}_{\mathrm{GIM}}}
R_{\mathrm{sum}}\left(\mathbf{F}(\mathbf{B})\right),$$
and
$$
R_{\mathrm{sum}}^{\star,\mathrm{PIM}}
\triangleq
\max_{\mathbf{B}\in\mathcal{B}_{\mathrm{PIM}}}
R_{\mathrm{sum}}\left(\mathbf{F}(\mathbf{B})\right).$$
Because the PIM feasible set is a subset of the GIM feasible set, the globally optimal sum-rate under GIM cannot be lower than that under PIM, i.e.,
\begin{equation}
R_{\mathrm{sum}}^{\star,\mathrm{GIM}}
\geq
R_{\mathrm{sum}}^{\star,\mathrm{PIM}}.
\end{equation}
\end{remark}

\begin{remark}[Comparison with fully digital precoding]
For fully digital precoding schemes, the precoding matrix $\mathbf{W}\in\mathbb{C}^{N_t\times K}$ is optimized directly in the baseband. The corresponding transmit signal is
\begin{equation}
\mathbf{x}_{\mathrm{FD}}
=
\sqrt{\frac{P_t}{K}}
\mathbf{W}\mathbf{s},
\end{equation}
where $\mathbf{W}$ satisfies the normalized transmit-power constraint
$
\mathrm{tr}\left(\mathbf{W}^{H}\mathbf{W}\right)\leq K$.
The fully digital benchmark is therefore obtained from maximizing the sum-rate
\begin{equation}
\label{EQ:PFD}
\begin{aligned}
\mathrm{(P\text{-}FD)}\quad
\max_{\mathbf{W}}\quad
R_{\mathrm{sum}}(\mathbf{W})
\qquad
\mathrm{s.t.}\quad
\mathrm{tr}
\left(
\mathbf{W}^{H}\mathbf{W}
\right)
\leq K.
\end{aligned}
\end{equation}
Unlike fully digital precoding, which is only constrained by the Frobenius-norm power budget, the MiLAC-induced precoder is physically generated through $\mathbf{B}\mapsto\boldsymbol{\Theta}(\mathbf{B})\mapsto\mathbf{F}(\mathbf{B})$ and satisfies $\mathbf{F}^{H}(\mathbf{B})\mathbf{F}(\mathbf{B})\preceq\mathbf{I}$. Consequently, the performance gap between MiLAC and fully digital precoding mainly originates from two coupled factors: the possible reduction in MiLAC transmission coefficient and the structural limitation on precoder design flexibility.
\end{remark}

\section{SVD-Based Feasible-Set Characterization and Search Algorithm}
\label{Section3}

In this section, we develop an SVD-based exhaustive search algorithm to solve the MiLAC precoding optimization problem (P1). Since the mapping
$
\mathbf{B}\mapsto\boldsymbol{\Theta}(\mathbf{B})\mapsto\mathbf{F}(\mathbf{B})
$
is highly nonlinear and non-convex, it is difficult to solve (P1) directly over the susceptance matrix $\mathbf{B}$.
In particular, the PIM constraint
$\boldsymbol{\Theta}_{11}(\mathbf{B})=\mathbf{0}$ defines a nonlinear lower-dimensional subset in the $\mathbf{B}$-space. To overcome this difficulty, we perform the search in the induced precoding matrix $\mathbf{F}(\mathbf{B})$ rather than directly over $\mathbf{B}$.

The key observation is that the difference between GIM and PIM is fully captured by the structure of
$\mathbf{F}(\mathbf{B})=\boldsymbol{\Theta}_{21}(\mathbf{B})$: under the GIM model,
$
\mathbf{F}^H(\mathbf{B})\mathbf{F}(\mathbf{B})\preceq\mathbf{I},
$
whereas under the PIM model,
$
\mathbf{F}^H(\mathbf{B})\mathbf{F}(\mathbf{B})=\mathbf{I}.
$
More generally, any precoding matrix $\mathbf{F}$ satisfying
$
\mathbf{F}^{H}\mathbf{F}\preceq\mathbf{I}
$
is MiLAC-admissible and can be embedded as the RF-to-antenna block of a larger unitary symmetric scattering matrix $\boldsymbol{\Theta}$, corresponding to a lossless reciprocal network~\cite{nerini2025capacity,wu2026microwavelinearanalogcomputer}. 
The connection back to the susceptance-domain realization follows from the inverse Cayley transform~\cite{pozar2011microwave,nerini2026physics}. If the completed scattering matrix satisfies that $\boldsymbol{\Theta}+\mathbf{I}$ is nonsingular, the corresponding susceptance matrix is given by
\begin{equation}
\label{EQ:InverseCayley}
\mathbf{B}
=
\frac{j}{Z_0}
\left(\boldsymbol{\Theta}-\mathbf{I}\right)
\left(\boldsymbol{\Theta}+\mathbf{I}\right)^{-1}.
\end{equation}
The resulting $\mathbf{B}$ is finite and real-symmetric.

\subsection{Parameterization via Complex Givens Rotations }

We first parameterize the MiLAC-induced precoding matrix by its SVD as
\begin{equation}
\label{EQ:SVD}
\mathbf{F}(\mathbf{B})
=
\mathbf{U}\boldsymbol{\Sigma}\mathbf{V}^H,
\end{equation}
where $\mathbf{U}\in\mathbb{C}^{N_t\times K}$ is semi-unitary with
$\mathbf{U}^H\mathbf{U}=\mathbf{I}_K$, $\mathbf{V}\in\mathbb{C}^{K\times K}$ is unitary with $\mathbf{V}^H\mathbf{V}=\mathbf{I}_K$, and
$
\boldsymbol{\Sigma}
=
\operatorname{diag}(s_1,\ldots,s_K)
$
with diagonal elements $s_1,\ldots,s_K$ denoting the singular values.

Thus, the feasible sets of GIM and PIM problems are distinguished by the singular values. For the GIM problem (P1) in~\eqref{EQ:P1}, the constraint
$\mathbf{F}(\mathbf{B})^H\mathbf{F}(\mathbf{B})\preceq\mathbf{I}$ is equivalent to
\begin{equation}
\label{EQ:GIM_Singular}
0\leq s_k\leq 1,
\qquad
k=1,\ldots, K.
\end{equation}

For the PIM special case, the condition
$\mathbf{F}(\mathbf{B})^H\mathbf{F}(\mathbf{B})=\mathbf{I}$ in ~\eqref{Condi:PIM} implies
\begin{equation}
\label{EQ:PIM_Singular}
s_1=s_2=\cdots=s_K=1, 
\end{equation}
reducing the PIM precoding matrix to
$
\label{EQ:PIM_F}
\mathbf{F}(\mathbf{B})
=
\mathbf{U}\mathbf{V}^H
$. In other words, PIM achieves a unit transmission coefficient, but it also removes the additional degrees of freedom associated with singular-value shaping.


Given a candidate precoding matrix $\mathbf{F}$ generated by~\eqref{EQ:SVD}, the corresponding sum-rate is obtained as
\begin{equation}
\label{EQ:RateF}
R_{\mathrm{sum}}(\mathbf{F})
=
\sum_{k=1}^{K}
\log_2
\left(
1+
\frac{
\frac{P_t}{K}
|\mathbf{h}_k^H\mathbf{f}_k|^2
}{
\frac{P_t}{K}
\sum_{i\neq k}
|\mathbf{h}_k^H\mathbf{f}_i|^2
+
\sigma_k^2
}
\right),
\end{equation}
where $\mathbf{f}_k$ denotes the $k$-th column of $\mathbf{F}$. Therefore, the GIM and PIM search problems can be rewritten as
\begin{equation}
\label{EQ:P1_SVD}
\begin{aligned}
\max_{\mathbf{U},\boldsymbol{\Sigma},\mathbf{V}}
\quad
&
R_{\mathrm{sum}}
\left(
\mathbf{U}\boldsymbol{\Sigma}\mathbf{V}^H
\right)
\\
\mathrm{s.t.}\quad
&
\mathbf{U}^H\mathbf{U}=\mathbf{I}_K,
\quad
\mathbf{V}^H\mathbf{V}=\mathbf{I}_K,
\\
&
0\leq s_k\leq 1,\quad k=1,\ldots,K,
\end{aligned}
\end{equation}
and
\begin{equation}
\label{EQ:P2_SVD}
\begin{aligned}
\max_{\mathbf{U},\mathbf{V}}
\quad
&
R_{\mathrm{sum}}
\left(
\mathbf{U}\mathbf{V}^H
\right)
\\
\mathrm{s.t.}\quad
&
\mathbf{U}^H\mathbf{U}=\mathbf{I}_K,
\quad
\mathbf{V}^H\mathbf{V}=\mathbf{I}_K.
\end{aligned}
\end{equation}

The above formulations provide a unified basis for searching over GIM or PIM precoder. The remaining challenge is how to parameterize and explore the unitary factors $\mathbf{U}$ and $\mathbf{V}$.

To address this challenge, the semi-unitary factor $\mathbf{U}$ is parameterized as a cascade of complex Givens rotations~\cite{golub2013matrix}:
\begin{equation}
\label{EQ:U_Givens}
\mathbf{U}
=
\left(
\prod_{\ell=1}^{P_U}
\mathbf{G}_{i_\ell j_\ell}(\theta_\ell,\phi_\ell)
\right)
\mathbf{U}_0,
\end{equation}
where $\mathbf{U}_0\in\mathbb{C}^{N_t\times K}$ is a reference semi-unitary matrix, and $\mathbf{G}_{i_\ell j_\ell}(\theta_\ell,\phi_\ell)\in\mathbb{C}^{N_t\times N_t}$ denotes a complex Givens rotation acting on the coordinate pair $(i_\ell,j_\ell)$. The parameters $\theta_\ell$ and $\phi_\ell$ denote the rotation angle and phase, respectively. Similarly, the unitary matrix $\mathbf{V}$ can be parameterized as
$\mathbf{V}
=
\prod_{\ell=1}^{P_V}
\widetilde{\mathbf{G}}_{i_\ell j_\ell}(\vartheta_\ell,\varphi_\ell)$,
where $\widetilde{\mathbf{G}}_{i_\ell j_\ell}(\vartheta_\ell,\varphi_\ell)\in\mathbb{C}^{K\times K}$ is a Givens rotation in the $K$-dimensional right-unitary space.

This parameterization automatically preserves the semi-unitary or unitary structure of the factors and avoids additional feasibility projection during optimization.

\begin{algorithm}[tbp]
\caption{SVD-based Search Algorithm for Solving (P1)}
\label{alg:unified}
\begin{algorithmic}[1]
\Procedure{}{}
    \State Represent precoding matrix $\mathbf{F(\mathbf{B})} = \mathbf{U} \boldsymbol{\Sigma} \mathbf{V}^H$.
    \State Discretize the diagonal entries $s_{k}$ of  $\boldsymbol{\Sigma}$ over $[0,1]$, as implied by the impedance matching condition.
    \If{$N_t$ and $K$ are small}
		\State Parameterize full unitary matrices $\mathbf{U}$ and $\mathbf{V}$.
		\State Exhaustive search all combinations of $(\mathbf{U}, \mathbf{\Sigma}, \mathbf{V})$ for the optimal solution.
    \Else
 \State Restrict the matrix to $\mathbf{F}(\mathbf{B}) = \mathbf{U}_{\text{base}} \widetilde{\mathbf{U}} \boldsymbol{\Sigma} \widetilde{\mathbf{V}}^{H}$.       
\State Perform SVD of channel matrix $\mathbf{H}$ and set: \quad $\mathbf{U}_{\text{base}} \leftarrow$ $K$ non-zero left singular vectors.
        \State Parameterize unitary matrices\quad $\widetilde{\mathbf{U}}$ and $\widetilde{\mathbf{V}}$ via reduced Givens rotations.
          \State  Initialize the reduced search parameters.
\Repeat 
    \State Update $\widetilde{\mathbf{U}}$, $\mathbf{\Sigma}$, and $\widetilde{\mathbf{V}}$ alternately.
\Until{convergence is reached} 
    \EndIf
    \State \textbf{return} optimized $\mathbf{F}(\mathbf{B})$ and corresponding $\mathbf{B}$.
\EndProcedure
\end{algorithmic}
\end{algorithm}

\subsection{Exhaustive Search for Small/Medium Size Systems}
For small or medium size systems, the complete SVD parameter space can be discretized and exhaustively searched. For example, when $N_t=K=2$, any feasible matrix can be expressed as
\begin{equation}
\mathbf{F}
=
\mathbf{U}
\begin{bmatrix}
s_1 & 0\\
0 & s_2
\end{bmatrix}
\mathbf{V}^H,
\end{equation}
where $\mathbf{U}$ and $\mathbf{V}$ are $2\times2$ unitary matrices. A complex Givens rotation is given by
$\mathbf{G}(\theta,\tilde{\beta})
=
\begin{bmatrix}
\cos\theta & e^{j\tilde{\beta}}\sin\theta\\
-e^{-j\tilde{\beta}}\sin\theta & \cos\theta
\end{bmatrix}$,
where $\theta\in[0,\pi/2]$ and $\tilde{\beta}\in[0,2\pi)$. By introducing $\mathbf{G}(\theta,\tilde{\beta})$, any $2\times2$ unitary matrix can be written as
\begin{equation}
\label{EQ:U2}
\mathbf{U}(\phi,\alpha,\beta,\theta)
=
e^{j\phi}
\begin{bmatrix}
e^{j\alpha}\cos\theta & e^{j\beta}\sin\theta\\
-e^{-j\beta}\sin\theta & e^{-j\alpha}\cos\theta
\end{bmatrix},
\end{equation}
where $\beta = \tilde{\beta} - \alpha$, $\phi,\alpha,\beta\in[0,2\pi)$ and $\theta\in[0,\pi/2]$. See Appendix A for more information. The same parameterization can be applied to $\mathbf{V}$. The phase grids can be constructed as
$
\phi,\alpha,\beta
\in
\left\{
0,\frac{2\pi}{M},
\ldots,\frac{(M-1)2\pi}{M}
\right\},
$ 
and the rotation-angle grid is given by
$
\theta
\in
\left\{
0,\frac{\pi/2}{T-1},
\ldots,
\frac{\pi}{2}
\right\}$, where $M$ and $T$ denote the grid resolutions.

For GIM, the singular values $s_1$ and $s_2$ are discretized over $[0,1]$, whereas for PIM they are fixed to $s_1=s_2=1$. Each candidate precoder is then constructed and evaluated. 
The number of evaluations for full-dimensional exhaustive search grows exponentially with the number of real-valued search parameters, which can be approximated as
$$
\label{EQ:FullComplexity}
\mathcal{O}
\left(
(L_\theta L_\phi)^{D_U+D_V}
L_s^K
\right),
$$
where $L_\theta$ and $L_\phi$ denote the numbers of grid points for rotation angles and phases, respectively, and $L_s$ denotes the number of grid points for each singular value. The quantities
$
D_U=2N_tK-K^2,
D_V=K^2
$
represent the real degrees of freedom of the matrix $\mathbf{U}$ and $\mathbf{V}$, respectively. For PIM, the factor $L_s^K$ is removed because all the singular values are one. 

Note that the proposed design with full-dimensional SVD-based exhaustive search is applicable to general system dimensions. However, it becomes computationally prohibitive for large-size systems as the numbers of transmit antennas and users become large.

\subsection{Channel-Guided Acceleration of SVD-Based Search}

The exhaustive search process can be accelerated by restricting the left singular space to a channel‑guided subspace and updating the remaining SVD variables in a block‑coordinate manner. 

Denoting the SVD of the aggregate channel matrix as 
$
\mathbf{H}
=
\mathbf{U}_H
\boldsymbol{\Sigma}_H
\mathbf{V}_H^H, 
$
we construct a basis
$
\mathbf{U}_{\mathrm{base}}
=
[\mathbf{u}_{H,1},\ldots,\mathbf{u}_{H,K}]
\in\mathbb{C}^{N_t\times K},
$
where $\mathbf{u}_{H,1},\ldots,\mathbf{u}_{H,K}$ are the $K$ non-zero left singular vectors of $\mathbf{H}$. The precoding matrix is then parameterized as
\begin{equation}
\label{EQ:ReducedF}
\mathbf{F}
=
\mathbf{U}_{\mathrm{base}}
\widetilde{\mathbf{U}}
\boldsymbol{\Sigma}
\widetilde{\mathbf{V}}^H,
\end{equation}
where $\widetilde{\mathbf{U}}\in\mathbb{C}^{K\times K}$, $\widetilde{\mathbf{V}}\in\mathbb{C}^{K\times K}$, and $\boldsymbol{\Sigma}
=
\operatorname{diag}(s_1,\ldots,s_K)$ satisfies the same singular-value constraints as in~\eqref{EQ:GIM_Singular} or~\eqref{EQ:PIM_Singular}. 
The effective left factor
$\mathbf{U}_{\mathrm{base}}\widetilde{\mathbf{U}}$
remains semi-unitary.

The reduced factors $\widetilde{\mathbf{U}}$ and $\widetilde{\mathbf{V}}$ are parameterized by a limited number of complex Givens rotations as
$
\widetilde{\mathbf{U}}
=
\prod_{\ell=1}^{P_U}
\mathbf{G}_{i_\ell j_\ell}(\theta_\ell,\phi_\ell),
\qquad
\widetilde{\mathbf{V}}
=
\prod_{\ell=1}^{P_V}
\widetilde{\mathbf{G}}_{i_\ell j_\ell}(\vartheta_\ell,\varphi_\ell),
$
where $P_U$ and $P_V$ control the number of retained Givens rotations for the left and right factors, respectively. 
The search complexity can be further reduced via block-coordinate updates, which sequentially optimize $\widetilde{\mathbf{U}}$, $\boldsymbol{\Sigma}$, and $\widetilde{\mathbf{V}}$ instead of jointly enumerating all parameter combinations. 
Since each retained Givens rotation is discretized with $L_\theta$ grid points for the rotation angle and $L_\phi$ grid points for the phase, the complexity per block-coordinate iteration becomes
$
\mathcal{O}
\left(
P_U L_\theta L_\phi
+
P_V L_\theta L_\phi
+
K L_s
\right),
$
which scales linearly with $P_U$, $P_V$, and $K$. 
This complexity reduction, however, may incur performance loss, since the resulting solution depends on the initialization and may be trapped in a local optimum. 
Moreover, as the dimensionality of the search parameter space increases, identifying high-quality solutions under a fixed grid resolution and computational budget becomes difficult.

The SVD-based search reveals the structural difference between GIM and PIM and can provide performance reference for small or medium size systems. The overall procedure of the proposed SVD-based search framework is summarized in Algorithm~\ref{alg:unified}.

\section{Unified Algorithm Via Projected WMMSE for Arbitrary Size Systems}
\label{Section4}

In this section, we develop a projected WMMSE-based algorithm for optimizing the MiLAC-induced precoding matrix $\mathbf{F}$, which has a much lower complexity than the SVD-based search framework and thus is more suitable for large-size systems.
As established in Section~\ref{Section2}, the feasible structure of $\mathbf{F}$ depends on the impedance-matching model: under GIM, $\mathbf{F}$ satisfies the contraction constraint
$
\mathbf{F}^{H}\mathbf{F}\preceq \mathbf{I}_{K},
$
whereas under PIM, it is restricted to the semi-unitary constraint
$
\mathbf{F}^{H}\mathbf{F}= \mathbf{I}_{K}.
$
Based on them, we first derive an efficient projected WMMSE update for GIM, where the projection is implemented by applying SVD to the tentative precoder followed by singular-value clipping. 
The PIM case can then be handled using the same framework by applying the corresponding semi-unitary projection.

Based on~\eqref{EQ:ReceivedSignalGeneric},~\eqref{EQ:SumRateGeneric} and~\eqref{EQ:FContraction}, the GIM precoding optimization problem is written as
\begin{equation}
\label{P:GIM_F_WMMSE}
\begin{aligned}
\max_{\mathbf{F}\in\mathbb{C}^{N_t\times K}}\quad
& R_{\mathrm{sum}}(\mathbf{F})\\
\mathrm{s.t.}\quad
& \mathbf{F}^{H}\mathbf{F}\preceq\mathbf{I}_{K}.
\end{aligned}
\end{equation}
Problem~\eqref{P:GIM_F_WMMSE} is non-convex due to the coupled multiuser interference terms in the sum-rate objective. To obtain an efficient iterative solution, we adopt the WMMSE reformulation~\cite{shi2011iteratively}.  
The basic idea is to introduce, for each user, a scalar receive equalizer $u_k$ and a positive MSE weight $q_k$, such that the original sum-rate maximization can be equivalently handled through a block-wise minimization over $\{u_k\}$, $\{q_k\}$, and the induced MiLAC precoder $\mathbf{F}$.

\subsection{WMMSE Reformulation and Block-Wise Updates for GIM}
We first establish the rate-WMMSE equivalence that connects the original sum-rate maximization problem with its WMMSE minimization reformulation.
Let $u_k$ denote the scalar receive equalizer for user $k$, and define the symbol estimate as
\begin{equation}
\hat{s}_k=u_k^{*}y_k .
\label{EQ:GIM_symbol_estimate}
\end{equation}
The corresponding MSE is given by
\begin{equation}
\begin{aligned}
e_k
=&
\mathbb{E}\left[|u_k^{*}y_k-s_k|^2\right] \\
=&
1
-
2\operatorname{Re}
\left\{
u_k^{*}
\sqrt{\frac{P_t}{K}}
\mathbf{h}_{k}^{H}\mathbf{f}_{k}
\right\}
+
|u_k|^2
\left(
\frac{P_t}{K}
\sum_{j=1}^{K}
|\mathbf{h}_{k}^{H}\mathbf{f}_{j}|^2
+
\sigma_k^2
\right).
\end{aligned}
\label{EQ:GIM_MSE}
\end{equation}
By introducing a positive MSE weight $q_k>0$ for each user, the original sum-rate maximization problem in~\eqref{P:GIM_F_WMMSE} can be equivalently transformed into a WMMSE minimization problem~\cite{shi2011iteratively}.
The rate-WMMSE equivalence is summarized in the following lemma.
\begin{lemma}[]
\label{lem:WMMSE_equivalence}
For any fixed feasible precoder $\mathbf{F}$, the achievable rate of user $k$ satisfies
\begin{equation}
\log_2\left(1+\gamma_k(\mathbf{F})\right)
=
\frac{1}{\log 2}
\left[
1
-
\min_{u_k,q_k>0}
\left(
q_k e_k-\log q_k
\right)
\right],
\label{EQ:rate_WMMSE_equiv}
\end{equation}
where $e_k$ is defined in~\eqref{EQ:GIM_MSE}. 
The minimum of the problem in~\eqref{EQ:rate_WMMSE_equiv} is achieved by
\begin{equation}
u_k^{\star}
=
\frac{
\sqrt{\frac{P_t}{K}}\mathbf{h}_{k}^{H}\mathbf{f}_{k}
}{
\frac{P_t}{K}
\sum_{j=1}^{K}
|\mathbf{h}_{k}^{H}\mathbf{f}_{j}|^2
+
\sigma_k^2
},
\label{EQ:GIM_u_update}
\end{equation}
and
\begin{equation}
q_k^{\star}
=
\left(e_k^{\star}\right)^{-1}.
\label{EQ:GIM_q_update}
\end{equation}
Here, $e_k^{\star}$ is the minimum MSE obtained with $u_k=u_k^{\star}$.
\end{lemma}
\begin{solidproof}
Please refer to Appendix B.
\end{solidproof}
According to Lemma~\ref{lem:WMMSE_equivalence}, maximizing the sum-rate over $\mathbf{F}$ is equivalent, up to a constant scaling factor, to minimizing the objective of the following problem 
\begin{equation}
\begin{aligned}
\min_{\mathbf{F},\,\{u_k\},\,\{q_k>0\}}
\quad
&
\sum_{k=1}^{K}
\left(
q_k e_k-\log q_k
\right) \\
\mathrm{s.t.}
\quad
&
\mathbf{F}^{H}\mathbf{F}\preceq \mathbf{I}_{K}.
\end{aligned}
\label{P:GIM_WMMSE}
\end{equation}
This naturally leads to an alternating optimization procedure. For a given precoder $\mathbf{F}$, the receive equalizers and MSE weights are updated in closed form according to~\eqref{EQ:GIM_u_update} and~\eqref{EQ:GIM_q_update}.

On the other hand, for fixed $\{u_k,q_k\}_{k=1}^{K}$, the $\mathbf{F}$-dependent part of the WMMSE objective can be written as the quadratic form
\begin{equation}
\operatorname{tr}\left(\mathbf{F}^{H}\mathbf{A}\mathbf{F}\right)
-
2\operatorname{Re}
\left\{
\operatorname{tr}\left(\mathbf{F}^{H}\mathbf{C}\right)
\right\},
\label{EQ:GIM_quadratic_F}
\end{equation}
where
\begin{equation}
\mathbf{A}
=
\sum_{k=1}^{K}
q_k|u_k|^2
\frac{P_t}{K}
\mathbf{h}_{k}\mathbf{h}_{k}^{H},
\label{EQ:GIM_A}
\end{equation}
and
\begin{equation}
\mathbf{C}
=
\left[
q_1\sqrt{\frac{P_t}{K}}u_1\mathbf{h}_1,
\ldots,
q_K\sqrt{\frac{P_t}{K}}u_K\mathbf{h}_K
\right].
\label{EQ:GIM_C}
\end{equation}
Therefore, under the GIM model, the fixed-$\{u_k,q_k\}$ $\mathbf{F}$-subproblem becomes
\begin{equation}
\begin{aligned}
\min_{\mathbf{F}}\quad
&
\operatorname{tr}\left(\mathbf{F}^{H}\mathbf{A}\mathbf{F}\right)
-
2\operatorname{Re}
\left\{
\operatorname{tr}\left(\mathbf{F}^{H}\mathbf{C}\right)
\right\} \\
\mathrm{s.t.}\quad
&
\mathbf{F}^{H}\mathbf{F}\preceq\mathbf{I}_{K}.
\end{aligned}
\label{P:GIM_F_subproblem}
\end{equation}

The subproblem in~\eqref{P:GIM_F_subproblem} is a convex quadratic problem over the contraction set $\mathbf{F}^{H}\mathbf{F}\preceq\mathbf{I}_{K}$, and thus can in principle be solved by existing convex solvers such as CVX~\cite{cvx}. 
However, solving such a constrained quadratic program at each WMMSE iteration would introduce considerable computational overhead. 
Moreover, a solver-based treatment does not directly provide a unified implementation for PIM, whose feasible set is the non-convex semi-unitary set $\mathbf{F}^{H}\mathbf{F}=\mathbf{I}_{K}$. 
Therefore, we adopt a projected-gradient update in this work, for which the projections onto the GIM or PIM MiLAC-induced feasible sets admit closed-form expressions.

Specifically, the Wirtinger gradient of~\eqref{EQ:GIM_quadratic_F} with respect to the complex conjugate $\mathbf{F}^{*}$ is given by
\begin{equation}
\mathbf{G}_{F}
=
\mathbf{A}\mathbf{F}-\mathbf{C}.
\label{EQ:GIM_F_gradient}
\end{equation}
Given a stepsize $\mu>0$, the tentative precoder matrix is
\begin{equation}
\mathbf{Z}
=
\mathbf{F}
-
\mu
\left(
\mathbf{A}\mathbf{F}-\mathbf{C}
\right).
\label{EQ:GIM_tentative_update}
\end{equation}
Then, the updated precoding matrix can be obtained by projecting $\mathbf{Z}$ back onto the GIM feasible set
\begin{equation}
\mathcal{F}_{\mathrm{GIM}}
=
\left\{
\mathbf{F}:
\mathbf{F}^{H}\mathbf{F}\preceq\mathbf{I}_{K}
\right\}.
\label{EQ:GIM_feasible_set}
\end{equation}

The key step is to compute the Euclidean projection of the tentative precoder $\mathbf{Z}$ onto the GIM feasible set. 
Since the GIM constraint
$
\mathbf{F}^{H}\mathbf{F}\preceq\mathbf{I}_{K}
$
is equivalent to $\|\mathbf{F}\|_2\leq 1$, the feasible set is a spectral-norm ball. 
Therefore, the projection can be performed in closed form by clipping the singular values of $\mathbf{Z}$.
Let the SVD of $\mathbf{Z}$ be
\begin{equation}
\mathbf{Z}
=
\mathbf{U}_{Z}
\operatorname{diag}(z_1,\ldots,z_K)
\mathbf{V}_{Z}^{H},
\label{EQ:Z_SVD}
\end{equation}
where $z_k\geq0$ denotes the $k$-th singular value. 
The Euclidean projection of $\mathbf{Z}$ onto $\mathcal{F}_{\mathrm{GIM}}$ is then obtained by
\begin{equation}
\Pi_{\mathrm{GIM}}(\mathbf{Z})
=
\mathbf{U}_{Z}
\operatorname{diag}
\left(
\min\{z_1,1\},\ldots,\min\{z_K,1\}
\right)
\mathbf{V}_{Z}^{H}.
\label{EQ:GIM_projection}
\end{equation}

This projection preserves the left and right singular vectors of $\mathbf{Z}$ and truncates only the singular values that exceed the unit threshold. 
At each iteration, for fixed $\{u_k,q_k\}$, the projected WMMSE precoder update under the GIM model is given by
\begin{equation}
\mathbf{F}^{+}
=
\Pi_{\mathrm{GIM}}
\left(
\mathbf{F}
-
\mu
\left(
\mathbf{A}\mathbf{F}-\mathbf{C}
\right)
\right).
\label{EQ:GIM_projected_update}
\end{equation}
A backtracking line search is then used to select $\mu$ such that the WMMSE objective with fixed $\{u_k,q_k\}$ is non-increasing after each projected update. The overall algorithm is summarized in Algorithm~\ref{alg:GIM_MiLAC_WMMSE}.

\subsection{Unified Projection for PIM}
We next demonstrate that the projected WMMSE-based algorithm can be readily applied to the PIM case. 
The updates of the receive equalizers $\{u_k\}$ and MSE weights $\{q_k\}$, as well as the gradient direction $\mathbf{G}_F$, remain the same as in the GIM case. 
The only required modification is the projection operation, which under the PIM model is performed over the semi-unitary feasible set
\begin{equation}
\mathcal{F}_{\mathrm{PIM}}
=
\left\{
\mathbf{F}:
\mathbf{F}^{H}\mathbf{F}=\mathbf{I}_{K} 
\right\}
\subseteq
\mathcal{F}_{\mathrm{GIM}}.
\label{EQ:PIM_feasible_set}
\end{equation}
Using the SVD of $\mathbf{Z}$ as in \eqref{EQ:Z_SVD}, the projection onto $\mathcal{F}_{\mathrm{PIM}}$ is
\begin{equation}
\Pi_{\mathrm{PIM}}(\mathbf{Z})
=
\mathbf{U}_{Z}\mathbf{V}_{Z}^{H}.
\label{EQ:PIM_projection}
\end{equation}
Thus, the PIM projection forces all singular values of the induced precoder to be one, yielding
\begin{equation}
\eta_{\mathrm{tx}}
=
\frac{\|\mathbf{F}\|_F^2}{K}
=
\frac{1}{K}
\sum_{k=1}^{K}
s_k^2(\mathbf{F})
=
1.
\end{equation}
In contrast, the GIM projection in~\eqref{EQ:GIM_projection} only clips singular values exceeding one and allows $\eta_{\mathrm{tx}}\leq1$. 
This difference directly reflects the power-radiation-efficiency versus precoding-flexibility tradeoff: PIM preserves full RF-to-antenna transmission efficiency but restricts the precoder to a semi-unitary structure, whereas GIM permits singular-value attenuation and provides additional degrees of freedom for precoder design for interference mitigation.

Since the overall WMMSE problem remains non-convex, the converged solution may depend on the initialization. 
To improve robustness, we employ a multi-start strategy.
For each feasible initial precoder, the algorithm alternately updates $\{u_k\}$, $\{q_k\}$, and $\mathbf{F}$ until convergence. 
In each iteration, $\mathbf{A}$ and $\mathbf{C}$ are first constructed from the updated $\{u_k,q_k\}$. A tentative precoder is then obtained by a gradient step and projected onto the corresponding feasible set, as in~\eqref{EQ:GIM_projected_update}. 
The projection is set to $\Pi_{\mathrm{GIM}}$ in~\eqref{EQ:GIM_projection} for GIM and to $\Pi_{\mathrm{PIM}}$ in~\eqref{EQ:PIM_projection} for PIM. 
Among all converged solutions, the one with the largest original sum-rate is selected as the final output.


\begin{algorithm}[tbp]
\caption{Unified Projected WMMSE-based Algorithm for Arbitrary Size MiLAC precoder}
\label{alg:GIM_MiLAC_WMMSE}
\begin{algorithmic}[1]
\Require Channel matrix $\mathbf{H}$, transmit-power budget $P_t$, noise variance $\{\sigma_k^2\}$, and feasible initial set $\mathcal{F}_{\mathrm{init}}$.
\Ensure Optimized MiLAC precoding matrix $\mathbf{F}^{\star}$.
   \State Set $R_{\mathrm{best}}=-\infty$.
   \For{each initial point $\mathbf{F}^{(0)}\in\mathcal{F}_{\mathrm{init}}$}
      \State Set $t=0$.
      \Repeat
         \State Update $\{u_k\}$ according to~\eqref{EQ:GIM_u_update}.
         \State Update $\{q_k\}$ according to~\eqref{EQ:GIM_q_update}.
         \State Construct $\mathbf{A}$ and $\mathbf{C}$ according to~\eqref{EQ:GIM_A} and~\eqref{EQ:GIM_C}.
         \State Compute the gradient $\mathbf{G}_{F}=\mathbf{A}\mathbf{F}^{(t)}-\mathbf{C}$.
         \State Choose stepsize $\mu$ by backtracking line search.
         \State Update
         $
         \mathbf{F}^{(t+1)}
         =
         \Pi_{\{ \mathrm{GIM},\mathrm{PIM}\} }
         \left(
         \mathbf{F}^{(t)}-\mu\mathbf{G}_{F}
         \right)
         $
         using~\eqref{EQ:GIM_projection} or~\eqref{EQ:PIM_projection}.
         \State Set $t\leftarrow t+1$.
      \Until{convergence is reached}
      \State Evaluate $R_{\mathrm{sum}}(\mathbf{F}^{(t)})$.
      \If{$R_{\mathrm{sum}}(\mathbf{F}^{(t)})>R_{\mathrm{best}}$}
          \State $\mathbf{F}^{\star}\leftarrow \mathbf{F}^{(t)}$ and $R_{\mathrm{best}}\leftarrow R_{\mathrm{sum}}(\mathbf{F}^{(t)})$.
      \EndIf
   \EndFor
   \State \textbf{return} $\mathbf{F}^{\star}$.
\end{algorithmic}
\end{algorithm}

\subsection{Convergence and Complexity Analysis}

For each initialization, the proposed algorithm alternately updates the receive equalizers, MSE weights, and induced MiLAC precoder. 
For a fixed procoder $\mathbf{F}$, the updates of $\{u_k\}$ and $\{q_k\}$ are optimal in closed form. 
For fixed $\{u_k,q_k\}$, the GIM projected-gradient step with backtracking line search guarantees that the WMMSE objective is non-increasing after the precoder update. 
Since the WMMSE objective is lower bounded, the generated objective sequence converges. 
Due to the nonconvexity of the original sum-rate maximization problem, global optimality is not guaranteed. 
Nevertheless, under standard regularity conditions\footnote{The standard regularity conditions include the boundedness of the feasible set, smoothness of the WMMSE objective with respect to $\mathbf{F}$ for fixed $\{u_k,q_k\}$, and a stepsize rule satisfying sufficient descent.}, any limit point of the generated sequence satisfies a first-order stationarity condition of the corresponding projected WMMSE problem.

It is worth noting that the above convex-projection argument applies directly to GIM because, for fixed $\{u_k,q_k\}$, the GIM subproblem in~\eqref{P:GIM_F_subproblem} is convex over the contraction set
$
\mathcal{F}_{\mathrm{GIM}}
=
\{\mathbf{F}:\mathbf{F}^{H}\mathbf{F}\preceq\mathbf{I}_{K}\}.
$
Therefore, this subproblem can be globally solved in principle. 
By contrast, the PIM counterpart is defined over the non-convex semi-unitary set
$
\mathcal{F}_{\mathrm{PIM}}
=
\{\mathbf{F}:\mathbf{F}^{H}\mathbf{F}=\mathbf{I}_{K}\},
$
and thus the projected PIM update does not enjoy the same convex-projection guarantee, nor does it guarantee global optimality for the corresponding fixed-$\{u_k,q_k\}$ subproblem. 
This distinction concerns only the fixed-$\{u_k,q_k\}$ $\mathbf{F}$-update subproblem; while the overall alternating WMMSE procedure remains non-convex for both GIM and PIM models, and global optimality of the original sum-rate maximization problem is not guaranteed in general.

The computational complexity of each iteration mainly comes from three parts. First, updating the receive equalizers and MSE weights requires computing the effective channels $\mathbf{h}_{k}^{H}\mathbf{f}_{j}$ for all user pairs, resulting in complexity $\mathcal{O}(N_tK^2)$. Second, constructing the matrix $\mathbf{A}$ in~\eqref{EQ:GIM_A} requires summing $K$ rank-one matrices of size $N_t\times N_t$, with complexity $\mathcal{O}(N_t^2K)$. Third, the projection step requires the SVD of an $N_t\times K$ matrix. Since $N_t\geq K$ in the considered setup, its complexity is approximately $\mathcal{O}(N_tK^2)$. Therefore, excluding the additional cost of the backtracking line search, the per-iteration complexity is
\begin{equation}
\mathcal{O}
\left(
N_t^2K+N_tK^2
\right).
\end{equation}
Compared with the fully SVD-based exhaustive search, whose complexity grows exponentially with the system dimension and grid resolution, the projected WMMSE method scales polynomially with $N_t$ and $K$. Therefore, it provides an efficient solver for arbitrary size systems with either GIM- or PIM-based MiLAC precoding, while preserving the corresponding MiLAC-induced physical constraints.

\begin{figure}[tbp]
  \centering
  \subfigure[]{
  \label{F1}
  \includegraphics[width=0.82\linewidth]{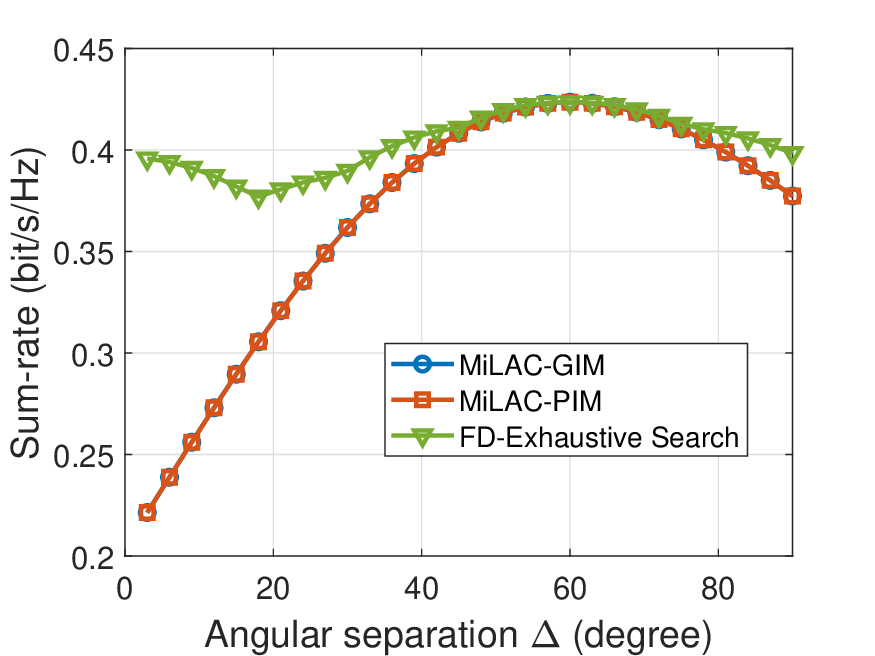}}
  \subfigure[]{
  \label{F2}
  \includegraphics[width=.82\linewidth]{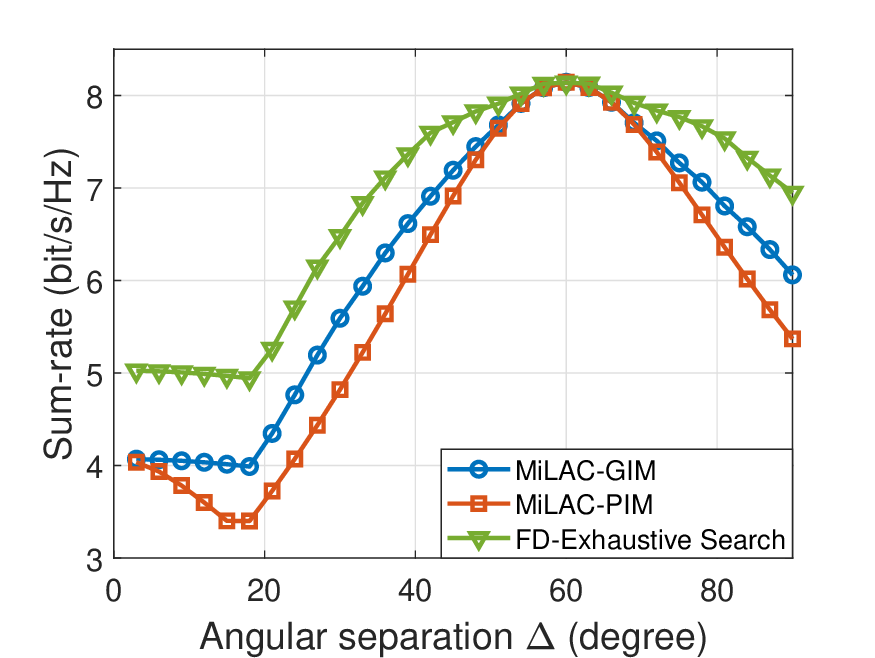}}
  \subfigure[]{
  \label{fig:los_eta}
  \includegraphics[width=.82\linewidth]{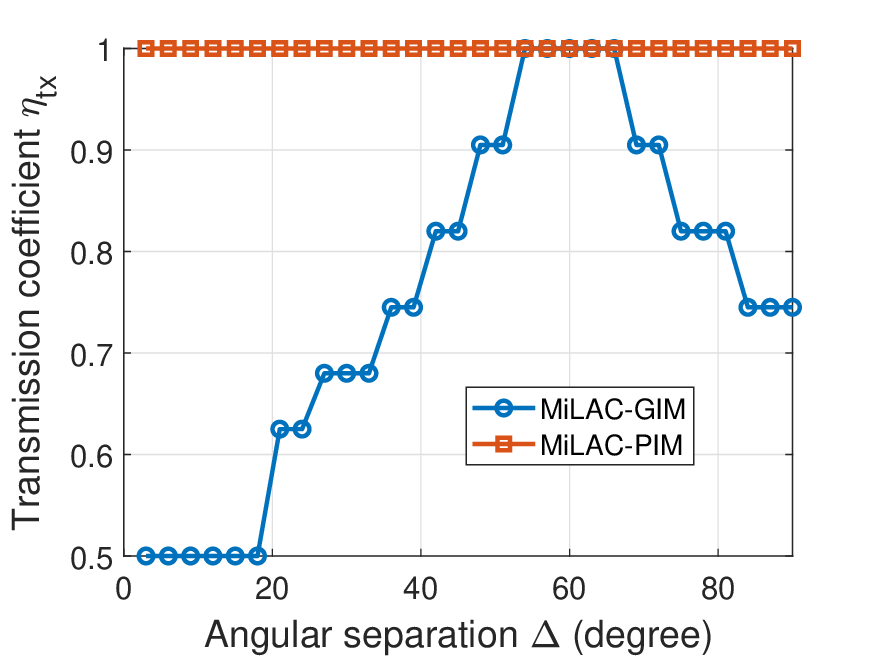}}
  \caption{Performance under the LoS channel. (a) Sum-rate, SNR = $-$5 dB. (b) Sum-rate, SNR = 15 dB. (c) Transmission coefficient, SNR = 15 dB.}
  \label{fig:los_sumrate}
\end{figure}

\section{Numerical Results}
\label{Section5}
We evaluate the proposed MiLAC precoding optimization framework under both line-of-sight (LoS) and Rayleigh fading channels. 
The simulations are designed to demonstrate two key aspects: the tradeoff between MiLAC radiated power and precoder design flexibility, and the performance gap between physically realizable MiLAC precoder and fully digital precoder.
Unless otherwise specified, the noise variance is normalized as $\sigma_k^2=1$ for all users, and the transmit-power budget is expressed in terms of the signal-to-noise ratio (SNR).

\subsection{Two-User LoS Channel}

We first consider a small-size two-user LoS channel with $N_t=K=2$ to examine how user channel correlation affects the performance of MiLAC with GIM or PIM configurations.
In this setup, the channel correlation between the two users is controlled by their angular separation, and the full-dimensional SVD-based exhaustive search is employed to exploit the feasible precoding structures under the GIM or PIM model.
For the GIM-based MiLAC search, the singular values are discretized as
$
s_k \in \{0,0.1,0.2,\ldots,1\}, \quad k=1,2,
$
whereas the PIM precoding matrix has unit singular values, i.e., $s_1=s_2=1$. 
For the fully digital benchmark, we also implement a grid-based exhaustive search over the SVD of the precoding matrix under the same total transmit-power constraint. 
Unlike MiLAC, the fully digital benchmark allows singular-value candidates satisfying $s_1^2+s_2^2\leq K$. 
The channel vector of user $k$ is modeled as
\begin{equation}
\mathbf{h}_k=\mathbf{a}_t(\theta_k),
\end{equation}
where
$
\mathbf{a}_t(\theta_k)
=
\frac{1}{\sqrt{N_t}}
\left[
1,
e^{j\pi\sin\theta_k},
\ldots,
e^{j\pi(N_t-1)\sin\theta_k}
\right]^T$ denotes the normalized transmit steering vector. The two users are symmetrically placed as
$
\theta_1=-\frac{\Delta}{2}, \theta_2=\frac{\Delta}{2}$,
where $\Delta$ denotes the angular separation between them. For the small-size system with $N_t=K=2$, the channel correlation is
$
\mathbf{h}_1^H\mathbf{h}_2
=
\frac{1}{2}
\left(
1+
e^{j\pi(\sin\theta_2-\sin\theta_1)}
\right)$. Strict channel orthogonality is achieved when
$
2\sin\left(\frac{\Delta}{2}\right)=1$, which gives $\Delta=60^\circ$.

Fig.~\ref{F1} and Fig.~\ref{F2} compare the sum-rate versus the angular separation $\Delta$ at SNR = $-$5 dB and SNR = 15 dB. At SNR = $-$5 dB, MiLAC-GIM and MiLAC-PIM achieve nearly identical performance, since the system is mainly noise-limited and interference mitigation is less critical. At SNR = 15 dB, MiLAC-GIM outperforms MiLAC-PIM over a wide range of angular separations, indicating that the enlarged feasible set of GIM is beneficial in interference-limited regimes due to more flexible interference shaping. 
When $\Delta$ approaches $60^\circ$, the two user channels become orthogonal, and the multiuser interference is naturally mitigated. 
In this case, the advantage of GIM over PIM diminishes, and the MiLAC-based schemes approach the fully digital benchmark.

Fig.~\ref{fig:los_eta} shows the corresponding MiLAC transmission coefficient performance at SNR =15 dB. As expected, MiLAC-PIM always achieves unit transmission coefficient, i.e., $\eta_{\mathrm{tx}}=1$, due to the zero-reflection constraint. 
In contrast, MiLAC-GIM generally has $\eta_{\mathrm{tx}}<1$ because non-zero RF-side reflection is allowed. 
Therefore, the sum-rate advantage of GIM at high transmit power is not attributed to a larger transmitted or radiated power. 
Instead, it results from the additional precoder design flexibility enabled by relaxing the PIM-induced semi-unitary constraint.
\begin{figure}[tbp]
  \centering
  \subfigure[]{
  \label{F3}
  \includegraphics[width=0.95\linewidth]{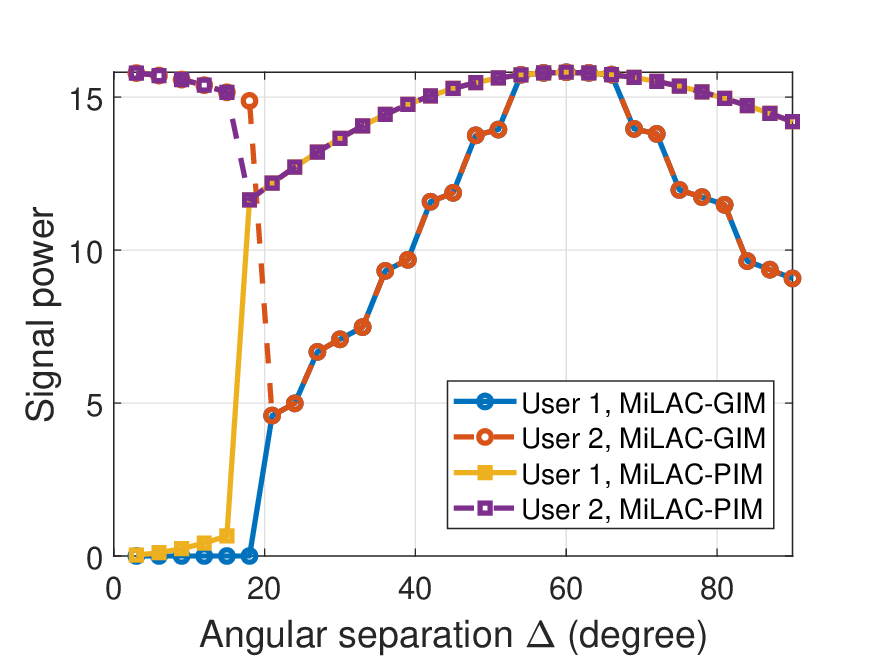}}
  \subfigure[]{
  \label{F4}
  \includegraphics[width=.95\linewidth]{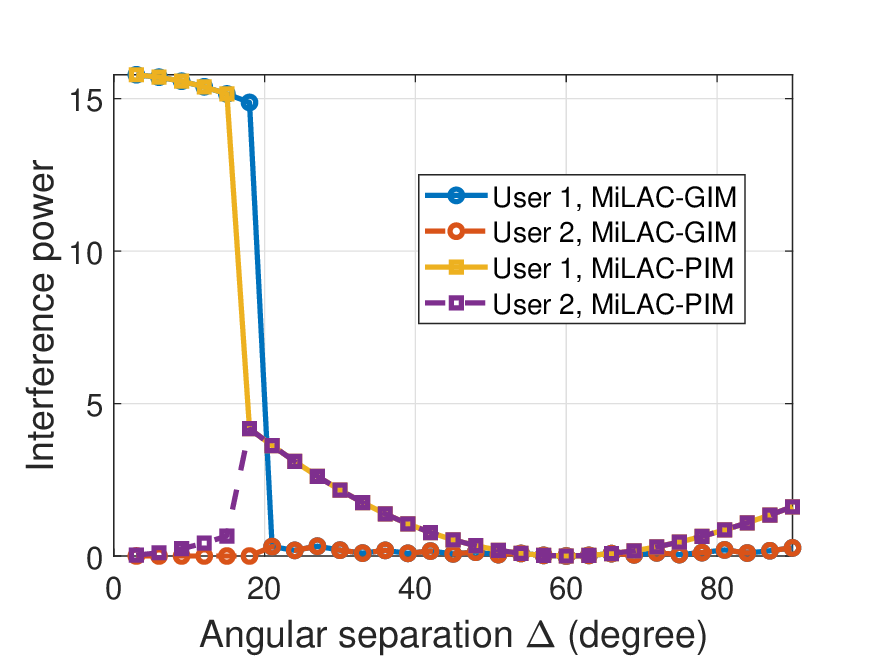}}
  \caption{Received signal power and interference power under the LoS channel. (a) Signal power. (b) Interference power.}
  \label{fig:los_power}
\end{figure}

\begin{figure}[tbp]
  \centering
  \subfigure[]{
  \label{Fig:Rayleigh_Nt4_sumrate}
  \includegraphics[width=0.95\linewidth]{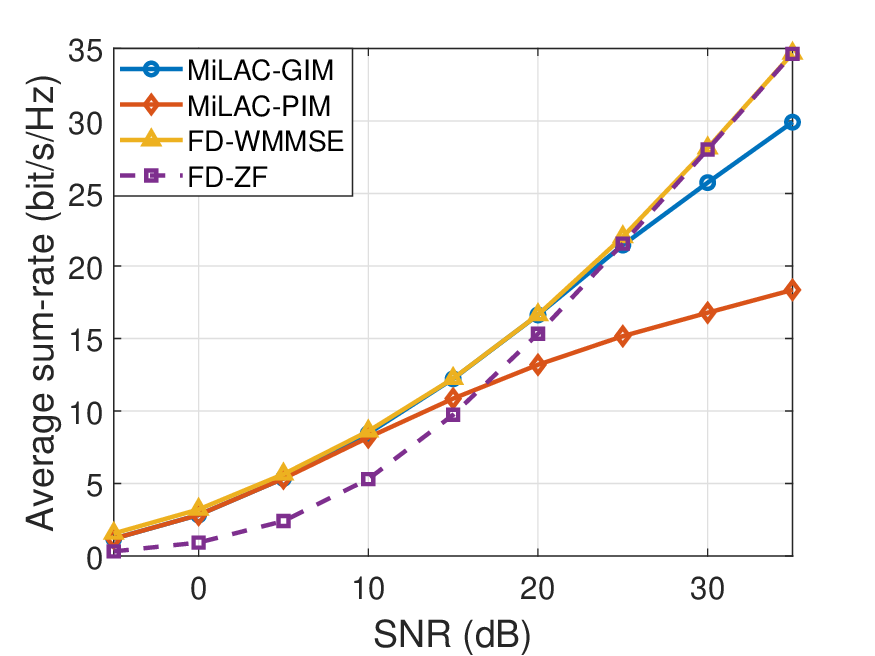}}
  \subfigure[]{
  \label{Fig:Rayleigh_Nt4_eta}
  \includegraphics[width=.95\linewidth]{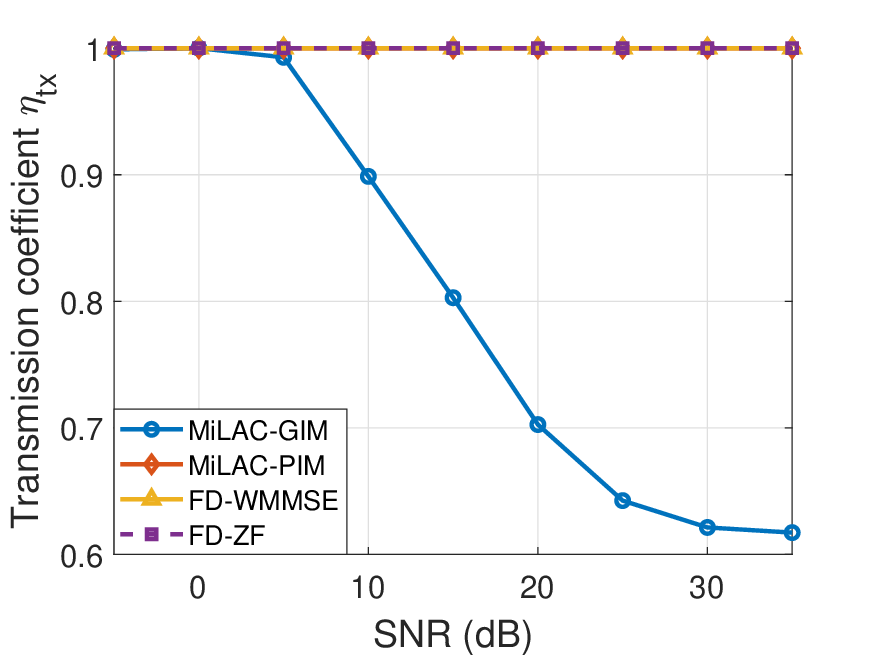}}
  \caption{Performance under Rayleigh fading channels with fixed $N_t=K=4$. (a) Sum-rate. (b) Transmission coefficient/normalized radiated power.}
  \label{Fig:Rayleigh_Nt4}
\end{figure}

To further identify the source of the GIM gain, we plot the desired signal power and interference power of the two users separately in Fig.~\ref{fig:los_power}. The results show that the advantage of MiLAC-GIM mainly comes from lower interference rather than stronger desired-signal power. 

\subsection{Multiuser Rayleigh Fading Channel}

We next evaluate the average sum-rate and MiLAC transmission coefficient under independent and identically distributed (i.i.d.) Rayleigh fading channels using the proposed projected WMMSE-based algorithm. 
In this set of simulations, we fix the number of users as $K=4$ and compare the performance under different numbers of transmit antennas and transmit SNR values. 
Two fully digital schemes are considered as benchmarks: fully digital weighted minimum mean-square error (FD-WMMSE)~\cite{shi2011iteratively} and fully digital zero-forcing (FD-ZF). 
FD-WMMSE serves as the main benchmark, where the receive equalizers, MSE weights, and digital precoder are iteratively updated under the normalized total-power constraint $\mathrm{tr}(\mathbf{W}^H\mathbf{W})\leq K$ specified in~\eqref{EQ:PFD}. FD-ZF is included as a suboptimal baseline, which is obtained by channel inversion followed by power normalization under the same constraint. 
The aggregate channel matrix $\mathbf{H}\in\mathbb{C}^{N_t\times K}$ is generated with independent CSCG entries, i.e.,
\begin{equation}
[\mathbf{H}]_{n,k}\sim \mathcal{CN}(0,1), 
\quad \forall n,k.
\end{equation}
For each transmit-power value and each $N_t$, the average sum-rate is computed over 100 independent channel realizations.

\begin{figure*}[tbp]
  \centering
\subfigure[]{
  \label{Fig:Rayleigh_Nt8_sumrate}
  \includegraphics[width=.47\linewidth]{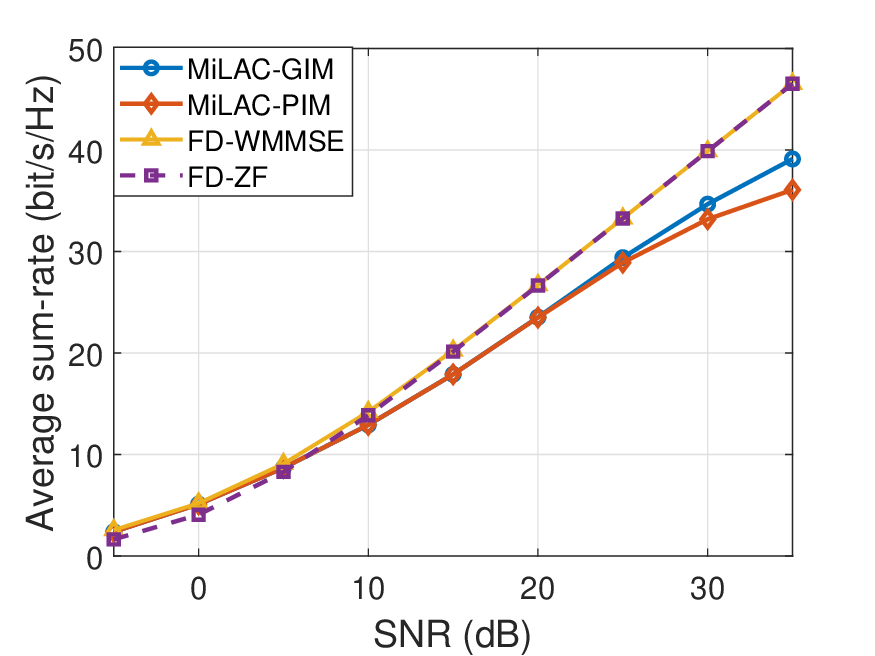}}
\subfigure[]{
  \label{Fig:Rayleigh_Nt8_eta}
  \includegraphics[width=.47\linewidth]{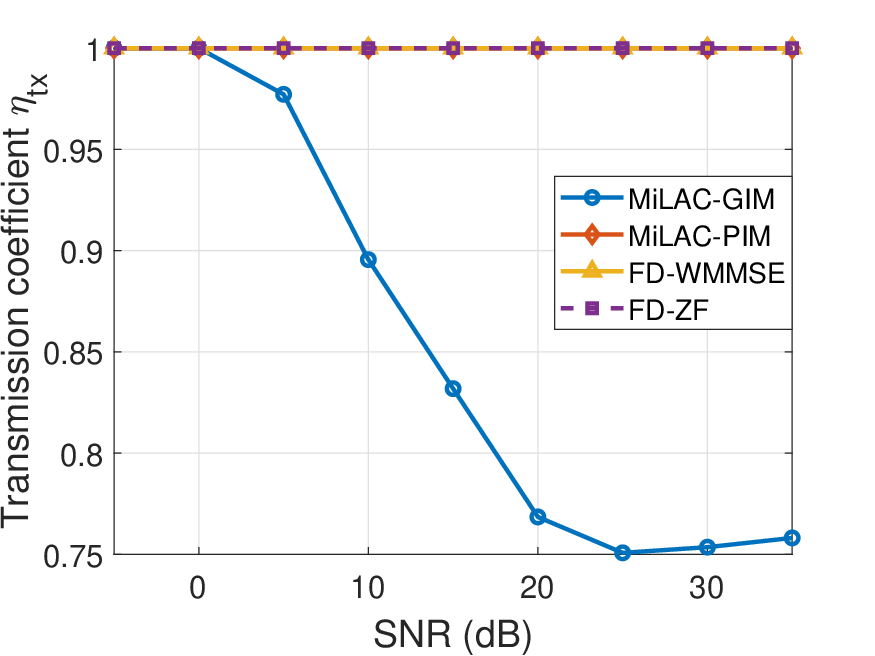}}
\subfigure[]{
  \label{Fig:Rayleigh_Nt32_sumrate}
  \includegraphics[width=.47\linewidth]{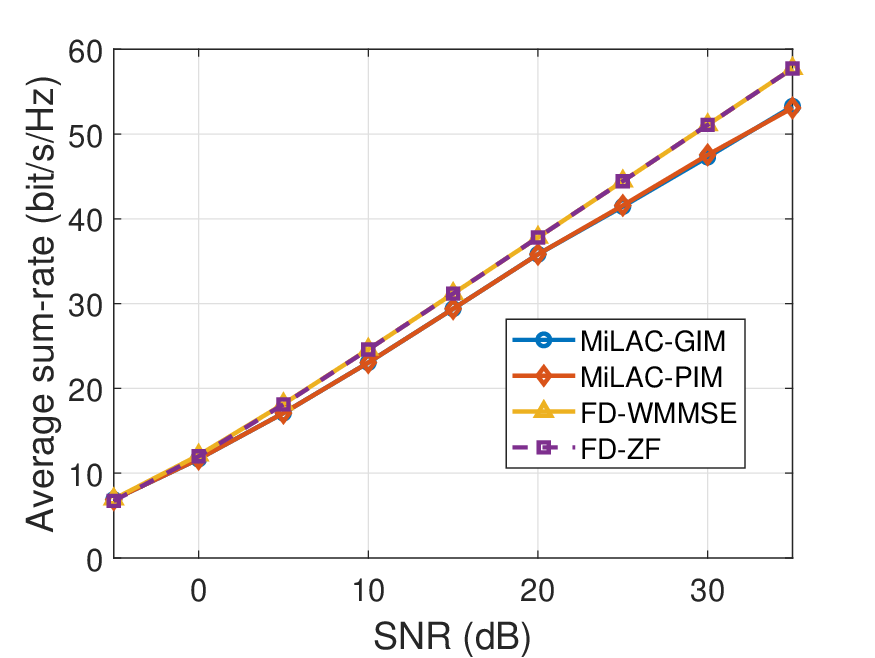}}
\subfigure[]{
  \label{Fig:Rayleigh_Nt32_eta}
  \includegraphics[width=.475\linewidth]{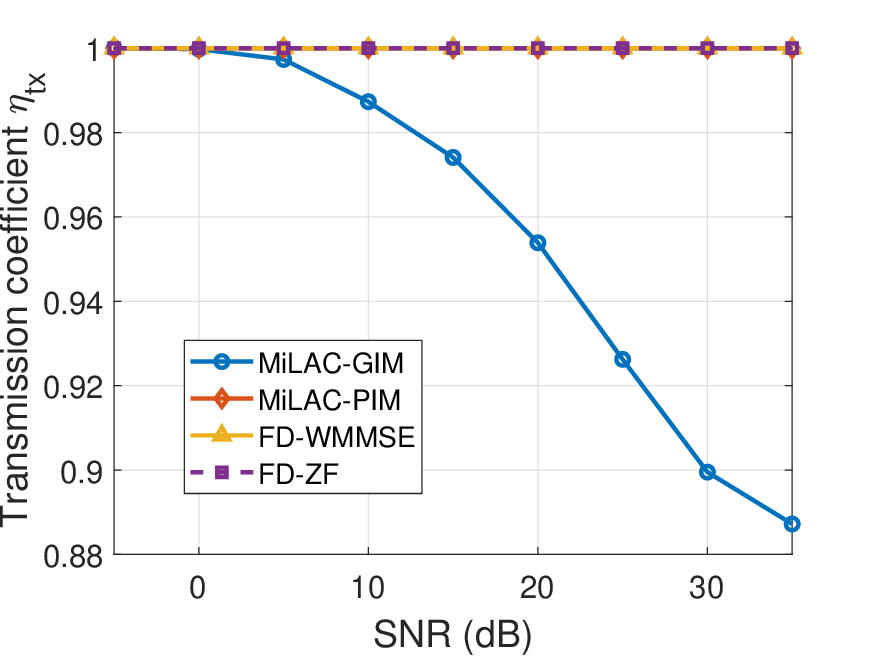}}
  \caption{Performance under Rayleigh fading channels with a fixed number of users $K=4$ and different numbers of transmit antennas $N_t$. (a) Sum-rate with $N_t=8$. (b) Transmission coefficient/normalized radiated power with $N_t=8$. (c) Sum-rate with $N_t=32$. (d) Transmission coefficient/normalized radiated power with $N_t=32$.}
  \label{Fig:Rayleigh_Nt_sumrate}
\end{figure*}

We first consider a highly loaded case with $N_t=K=4$, where the number of transmit antennas equals the number of served users and interference management becomes critical. As shown in Fig.~\ref{Fig:Rayleigh_Nt4_sumrate}, FD-WMMSE achieves the highest sum-rate due to its unconstrained baseband precoder design flexibility. In the low- and medium-SNR regimes, e.g., below approximately 15 dB, the MiLAC-based schemes can outperform FD-ZF, since ZF eliminates interference at the expense of desired-signal power and array gain, which is not always favorable when noise remains significant. 

Among the MiLAC-based schemes, MiLAC-GIM consistently outperforms MiLAC-PIM, and the performance gap becomes more evident as the SNR increases. 
This is because, as the system becomes interference-limited at high SNR, the additional singular-value degrees of freedom allowed by GIM become more important for interference mitigation.
By contrast, PIM enforces a semi-unitary precoding structure and always maintains $\eta_{\mathrm{tx}}=1$, which preserves full MiLAC transmission efficiency but limits the interference-shaping capability of the MiLAC precoder. 
The transmission-coefficient result in Fig.~\ref{Fig:Rayleigh_Nt4_eta} further supports this interpretation: MiLAC-GIM keeps $\eta_{\mathrm{tx}}\approx 1$ at low SNR, where preserving radiated power is more important, whereas at high SNR the optimized GIM solution exhibits a reduced $\eta_{\mathrm{tx}}$, reflecting the tradeoff between power radiation efficiency and interference mitigation flexibility.
Therefore, the sum-rate gain of GIM over PIM comes from the controlled degradation of transmission coefficient in exchange for improved interference mitigation.

To further examine the impact of the number of transmit antennas, we increase $N_t$ while keeping $K=4$ fixed. 
As shown in Fig.~\ref{Fig:Rayleigh_Nt_sumrate}, larger $N_t$ improves the sum-rate performance of all schemes and narrows the gap between MiLAC-based precoding and fully digital precoding. 
This is because a larger transmit array provides more spatial degrees of freedom and makes the user channels more separable, thereby mitigating multiuser interference. 
For $N_t=8$, MiLAC-GIM still achieves a visible gain over MiLAC-PIM in the high-SNR regimes, indicating that the additional singular-value degrees of freedom provided by GIM remain beneficial when the system is moderately loaded.

When the number of antennas further increases to $N_t=32$, MiLAC-GIM, MiLAC-PIM, FD-WMMSE, and FD-ZF exhibit much closer sum-rate performance. 
This suggests that, with sufficient transmit spatial degrees of freedom, the semi-unitary constraint imposed by PIM becomes less restrictive, since user channels tend to be nearly orthogonal and interference mitigation becomes easier. 
Consistently, as shown in Fig.~\ref{Fig:Rayleigh_Nt32_eta}, the transmission coefficient of MiLAC-GIM remains much closer to one than in the highly loaded $N_t=K=4$ case of Fig.~\ref{Fig:Rayleigh_Nt4_eta}. This indicates that, when the spatial dimension is sufficiently large, the MiLAC precoding optimization relies less on additional precoder design flexibility for interference mitigation and can thus maintain higher RF-to-antenna transmission efficiency.
These results indicate that the advantage of GIM is most pronounced in interference-limited multiuser scenarios, while PIM becomes nearly sufficient as the transmit array grows and the channel correlation decreases.

\section{Conclusion}
\label{Section6}

This paper studied physically realizable lossless reciprocal MiLAC-aided precoding optimization in downlink multiuser MISO communication systems. By explicitly linking the susceptance matrix, the scattering matrix, and the induced transmit precoding matrix, we established a physics-compliant framework for MiLAC precoding optimization under the PIM or GIM model. The analysis revealed that MiLAC performance is governed by two coupled factors: the MiLAC transmission coefficient and the available precoder design flexibility. PIM eliminates RF-side reflection and guarantees unit transmission coefficient, but restricts the induced precoder to a semi-unitary structure. In contrast, GIM relaxes the zero-reflection condition and enlarges the feasible precoding space, at the cost of possible radiated-power reduction. To characterize and optimize this tradeoff, we developed an SVD-based parametric search framework for feasible-set characterization and small or medium size systems performance benchmarking, and further proposed a unified algorithm based on projected WMMSE for GIM/PIM precoding optimization in arbitrary size systems. Numerical results under both LoS and Rayleigh fading channels showed that, in interference-limited multiuser scenarios, GIM outperforms PIM and achieves sum-rate performance close to that of the baseline fully digital precoding system, by sacrificing part of the radiated power to gain additional degrees of freedom for interference mitigation. These results indicate that imperfect impedance matching should not be treated merely as a hardware nonideality, but can be exploited as a useful design degree of freedom for MiLAC-aided multiuser precoding.

Future work beyond this paper may include extending the proposed GIM framework to wideband MiLAC-aided OFDM systems, where the frequency-dependent microwave response may introduce new tradeoffs among impedance matching, beam squint, and subcarrier-dependent precoding. 
Another promising direction is to develop transmission-coefficient-aware MiLAC design frameworks for extremely large-scale and/or movable antenna arrays.

\section*{Appendix A}
\label{app:details}

A complex Givens rotation acting on a $2\times2$ subspace is defined as
\begin{equation}
\mathbf{G}(\theta,\tilde{\beta})
=
\begin{bmatrix}
\cos\theta & e^{j\tilde{\beta}}\sin\theta \\
- e^{-j\tilde{\beta}}\sin\theta & \cos\theta
\end{bmatrix},
\label{eq:appendix_givens}
\end{equation}
where $\theta$ controls the rotation angle and $\tilde{\beta}$ determines the phase coupling between the two dimensions.

To connect this complex Givens rotation with the standard four-parameter representation of a general $2\times2$ unitary matrix, consider
\begin{equation}
\mathbf U = \mathbf G(\theta,\tilde{\beta}) \mathbf U_0,
\label{eq:appendix_U_general}
\end{equation}
where
\begin{equation}
\mathbf U_0
=
e^{j\phi}
\begin{bmatrix}
e^{j\alpha} & 0\\
0 & e^{-j\alpha}
\end{bmatrix}.
\label{eq:appendix_U0}
\end{equation}
Substituting \eqref{eq:appendix_U0} into \eqref{eq:appendix_U_general} gives
\begin{equation}
\mathbf U
=
e^{j\phi}
\mathbf G(\theta,\tilde{\beta})
\begin{bmatrix}
e^{j\alpha} & 0\\
0 & e^{-j\alpha}
\end{bmatrix}.
\end{equation}
By carrying out the matrix multiplication, we obtain
\begin{equation}
\mathbf U
=
e^{j\phi}
\begin{bmatrix}
e^{j\alpha}\cos\theta & e^{j(\tilde{\beta}-\alpha)}\sin\theta \\
- e^{j(\alpha-\tilde{\beta})}\sin\theta & e^{-j\alpha}\cos\theta
\end{bmatrix}.
\label{eq:appendix_intermediate}
\end{equation}
Defining
\begin{equation}
\beta = \tilde{\beta} - \alpha,
\label{eq:appendix_param_relation}
\end{equation}
we have $\alpha-\tilde{\beta}=-\beta$, and thus \eqref{eq:appendix_intermediate} can be rewritten as
\begin{equation}
\mathbf{U}(\phi,\alpha,\beta,\theta)
=
e^{j\phi}
\begin{bmatrix}
e^{j\alpha}\cos\theta & e^{j\beta}\sin\theta \\
- e^{-j\beta}\sin\theta & e^{-j\alpha}\cos\theta
\end{bmatrix}.
\label{eq:appendix_final}
\end{equation}
This is the standard four-parameter representation of a general $2\times2$ unitary matrix. 
Therefore, the complex Givens rotation parameterization used in the proposed SVD-based search is consistent with the conventional parameterization of $2\times2$ unitary matrices.

\section*{Appendix B: Proof of Lemma~\ref{lem:WMMSE_equivalence}}
The proof is motivated by~\cite{shi2011iteratively}. 
For fixed $\mathbf{F}$ and $q_k>0$, minimizing $q_ke_k-\log q_k$ with respect to $u_k$ is equivalent to minimizing the MSE $e_k$. The resulting MMSE receive equalizer is given by~~\eqref{EQ:GIM_u_update}, and the corresponding minimum MSE is
\begin{equation}
e_k^{\star}
=
\frac{1}{1+\gamma_k(\mathbf{F})}.
\end{equation}
For fixed $e_k$, minimizing $q_ke_k-\log q_k$ over $q_k>0$ gives $q_k^{\star}=1/e_k$ and
\begin{equation}
\min_{q_k>0}
\left(q_ke_k-\log q_k\right)
=
1+\log e_k.
\end{equation}
Substituting $e_k=e_k^{\star}=1/(1+\gamma_k(\mathbf{F}))$ gives
\begin{equation}
\min_{u_k,q_k>0}
\left(
q_ke_k-\log q_k
\right)
=
1-\log\left(1+\gamma_k(\mathbf{F})\right),
\end{equation}
or equivalently,
\begin{equation}
\log_2\left(1+\gamma_k(\mathbf{F})\right)
=
\frac{1}{\log 2}
\left[
1
-
\min_{u_k,q_k>0}
\left(
q_k e_k-\log q_k
\right)
\right].
\label{EQ:rate_WMMSE_equiv2}
\end{equation}

\bibliographystyle{IEEEtran}
\bibliography{Reference}

\begin{thebibliography}{10}
\providecommand{\url}[1]{#1}
\csname url@samestyle\endcsname
\providecommand{\newblock}{\relax}
\providecommand{\bibinfo}[2]{#2}
\providecommand{\BIBentrySTDinterwordspacing}{\spaceskip=0pt\relax}
\providecommand{\BIBentryALTinterwordstretchfactor}{4}
\providecommand{\BIBentryALTinterwordspacing}{\spaceskip=\fontdimen2\font plus
\BIBentryALTinterwordstretchfactor\fontdimen3\font minus
  \fontdimen4\font\relax}
\providecommand{\BIBforeignlanguage}[2]{{%
\expandafter\ifx\csname l@#1\endcsname\relax
\typeout{** WARNING: IEEEtran.bst: No hyphenation pattern has been}%
\typeout{** loaded for the language `#1'. Using the pattern for}%
\typeout{** the default language instead.}%
\else
\language=\csname l@#1\endcsname
\fi
#2}}
\providecommand{\BIBdecl}{\relax}
\BIBdecl

\bibitem{Emil2025enabling6g}
E.~Björnson, F.~Kara, N.~Kolomvakis, A.~Kosasih, P.~Ramezani, and M.~B.
  Salman, ``Enabling {6G} performance in the upper mid-band by transitioning
  from massive to gigantic {MIMO},'' \emph{IEEE Open J. Commun. Soc.}, vol.~6,
  pp. 5450--5463, Jun. 2025.

\bibitem{carvalho2020nonstationarities}
E.~D. Carvalho, A.~Ali, A.~Amiri, M.~Angjelichinoski, and R.~W. Heath,
  ``Non-stationarities in extra-large-scale massive {MIMO},'' \emph{IEEE
  Wireless Commun.}, vol.~27, no.~4, pp. 74--80, Aug. 2020.

\bibitem{xu2025distributed}
Y.~Xu, E.~G. Larsson, E.~A. Jorswieck, X.~Li, S.~Jin, and T.-H. Chang,
  ``Distributed signal processing for extremely large-scale antenna array
  systems: State-of-the-art and future directions,'' \emph{IEEE J. Sel. Topics
  Signal Process.}, vol.~19, no.~2, pp. 304--330, Mar. 2025.

\bibitem{larsson2014massive}
E.~G. Larsson, O.~Edfors, F.~Tufvesson, and T.~L. Marzetta, ``Massive {MIMO}
  for next generation wireless systems,'' \emph{IEEE Commun. Mag.}, vol.~52,
  no.~2, pp. 186--195, Feb. 2014.

\bibitem{emil2016massive}
E.~Björnson, E.~G. Larsson, and T.~L. Marzetta, ``Massive {MIMO}: ten myths
  and one critical question,'' \emph{IEEE Commun. Mag.}, vol.~54, no.~2, pp.
  114--123, Feb. 2016.

\bibitem{nerini2025analogI}
M.~Nerini and B.~Clerckx, ``Analog computing for signal processing and
  communications – part {I}: Computing with microwave networks,'' \emph{IEEE
  Trans. Signal Process.}, vol.~73, pp. 5183--5197, Dec. 2025.

\bibitem{nerini2025analog}
------, ``Analog computing for signal processing and communications—part
  {II}: Toward gigantic {MIMO} beamforming,'' \emph{IEEE Trans. Signal
  Process.}, vol.~73, pp. 5198--5212, Dec. 2025.

\bibitem{nerini2025capacity}
\BIBentryALTinterwordspacing
------, ``Capacity of {MIMO} systems aided by microwave linear analog computers
  ({MiLACs}),'' 2025. [Online]. Available:
  \url{https://arxiv.org/abs/2506.05983}
\BIBentrySTDinterwordspacing

\bibitem{nerini2025reduced}
------, ``{MIMO} systems aided by microwave linear analog computers:
  Capacity-achieving architectures with reduced circuit complexity,''
  \emph{IEEE Trans. Wireless Commun.}, vol.~25, pp. 14\,597--14\,610, Mar.
  2026.

\bibitem{foad2016hybrid}
F.~Sohrabi and W.~Yu, ``Hybrid digital and analog beamforming design for
  large-scale antenna arrays,'' \emph{IEEE J. Sel. Topics Signal Process.},
  vol.~10, no.~3, pp. 501--513, Apr. 2016.

\bibitem{wu2026microwavelinearanalogcomputer}
\BIBentryALTinterwordspacing
Z.~Wu, M.~Nerini, and B.~Clerckx, ``Microwave linear analog computer
  {(MiLAC)}-aided multiuser {MISO}: Fundamental limits and beamforming
  design,'' 2026. [Online]. Available: \url{https://arxiv.org/abs/2601.10060}
\BIBentrySTDinterwordspacing

\bibitem{zhou2026twolayer}
\BIBentryALTinterwordspacing
X.~Zhou, T.~Fang, Y.~Mao, and B.~Clerckx, ``Two-layer microwave linear analog
  computer {(MiLAC)}-aided multi-user {MISO} networks,'' 2026. [Online].
  Available: \url{https://arxiv.org/abs/2604.24303}
\BIBentrySTDinterwordspacing

\bibitem{peng2026hybriddigital}
\BIBentryALTinterwordspacing
Y.~Peng, Z.~Wu, and B.~Clerckx, ``Hybrid digital and microwave linear analog
  computer {(MiLAC)}-aided beamforming for multiuser {MIMO}-{OFDM} systems,''
  2026. [Online]. Available: \url{https://arxiv.org/abs/2604.26532}
\BIBentrySTDinterwordspacing

\bibitem{zhang2026channeles}
\BIBentryALTinterwordspacing
Q.~Zhang, M.~Nerini, and B.~Clerckx, ``Channel estimation in {MIMO} systems
  aided by microwave linear analog computers ({MiLACs}),'' 2026. [Online].
  Available: \url{https://arxiv.org/abs/2601.11438}
\BIBentrySTDinterwordspacing

\bibitem{nerini2026physics}
\BIBentryALTinterwordspacing
M.~Nerini and B.~Clerckx, ``Physics-compliant modeling and optimization of
  {MIMO} systems aided by microwave linear analog computers,'' 2026. [Online].
  Available: \url{https://arxiv.org/abs/2602.19379}
\BIBentrySTDinterwordspacing

\bibitem{fang2026performance}
\BIBentryALTinterwordspacing
T.~Fang, X.~Zhou, and Y.~Mao, ``On the performance of lossless reciprocal
  {MiLAC} architectures in multi-user networks,'' 2026. [Online]. Available:
  \url{https://arxiv.org/abs/2601.01834}
\BIBentrySTDinterwordspacing

\bibitem{fano1950theoretical}
R.~M. Fano, ``Theoretical limitations on the broadband matching of arbitrary
  impedances,'' \emph{J. Franklin Inst.}, vol. 249, no.~1, pp. 57--83, Jan.
  1950.

\bibitem{nemati2009design}
H.~M. Nemati, C.~Fager, U.~Gustavsson, R.~Jos, and H.~Zirath, ``Design of
  varactor-based tunable matching networks for dynamic load modulation of high
  power amplifiers,'' \emph{IEEE Trans. Microw. Theory Techn.}, vol.~57, no.~5,
  pp. 1110--1118, May 2009.

\bibitem{alibakhshikenari2021optimum}
M.~Alibakhshikenari \emph{et~al.}, ``Optimum power transfer in {RF} front end
  systems using adaptive impedance matching technique,'' \emph{Sci. Rep.},
  vol.~11, no.~1, p. 11825, Jun. 2021.

\bibitem{nerini2024universal}
M.~Nerini, S.~Shen, H.~Li, M.~Di~Renzo, and B.~Clerckx, ``A universal framework
  for multiport network analysis of reconfigurable intelligent surfaces,''
  \emph{IEEE Trans. Wireless Commun.}, vol.~23, no.~10, pp. 14\,575--14\,590,
  Oct. 2024.

\bibitem{pozar2011microwave}
D.~M. Pozar, \emph{Microwave engineering}.\hskip 1em plus 0.5em minus
  0.4em\relax John Wiley \& Sons, 2011.

\bibitem{golub2013matrix}
G.~H. Golub and C.~F. Van~Loan, \emph{Matrix Computations}.\hskip 1em plus
  0.5em minus 0.4em\relax Johns Hopkins Univ. Press, 2013.

\bibitem{shi2011iteratively}
Q.~Shi, M.~Razaviyayn, Z.-Q. Luo, and C.~He, ``An iteratively weighted {MMSE}
  approach to distributed sum-utility maximization for a {MIMO} interfering
  broadcast channel,'' \emph{IEEE Trans. Signal Process.}, vol.~59, no.~9, pp.
  4331--4340, Sep. 2011.

\bibitem{cvx}
\BIBentryALTinterwordspacing
M.~Grant and S.~Boyd, ``{CVX}: {MATLAB} software for disciplined convex
  programming,'' Mar. 2014. [Online]. Available: \url{http://cvxr.com/cvx/}
\BIBentrySTDinterwordspacing

\end{thebibliography}
%
%
%
%
%
%
%
%
%
%

\newpage

 



\end{document}